\documentclass[smallextended,envcountsect,envcountsame,envcountreset]{svjour2}

\smartqed  

\usepackage[english]{babel}
\usepackage[latin1]{inputenc}
\usepackage{indentfirst}
\usepackage{amsmath, amssymb, mathrsfs, dsfont, mathtools, empheq}
\usepackage{graphicx}
\usepackage[font={small,it}]{caption}
\usepackage{comment}

\newcommand{\eq}{\begin{equation}}
\newcommand{\eeq}{\end{equation}}
\newcommand{\bmat}{\begin{pmatrix}}
\newcommand{\emat}{\end{pmatrix}}
\newcommand{\virg}[1]{``#1''}

\newcommand{\R}{\mathbb R}

\newcommand{\Z}{\mathbb Z}
\newcommand{\C}{\mathbb C}

\newcommand{\1}{\mathds{1}\!}

\newcommand{\eps}{\varepsilon}
\renewcommand{\rho}{\varrho}

\newcommand{\al}{\alpha}
\newcommand{\ga}{\gamma}
\newcommand{\la}{\lambda}
\newcommand{\La}{\Lambda}
\newcommand{\D}{\mathscr D}
\newcommand{\h}{\textup h}					\renewcommand{\v}{\textup v}
\newcommand{\muh}{\mu_\h}					\newcommand{\muv}{\mu_\v}
\renewcommand{\l}{\textup l}				\renewcommand{\r}{\textup r}
\newcommand{\LE}{E_\l}						\newcommand{\RE}{E_\r}
\newcommand{\LB}{B_\l}						\newcommand{\RB}{B_\r}
\newcommand{\Il}{I_\l}						\newcommand{\Ir}{I_\r}
\newcommand{\xl}{x_\l}						\newcommand{\xr}{x_\r}
\newcommand{\Ixl}{I_{\l,\xl}}				\newcommand{\Ixr}{I_{\r,\xr}}
\newcommand{\Bxl}{B_{\l,\xl}}				\newcommand{\Bxr}{B_{\r,\xr}}
\newcommand{\bxl}{b_{\l,\xl}}				\newcommand{\bxr}{b_{\r,\xr}}
\newcommand{\epsxl}{\eps_{\l,\xl}}			\newcommand{\epsxr}{\eps_{\r,\xr}}
\newcommand{\etaxl}{\eta_{\l,\xl}}			\newcommand{\etaxr}{\eta_{\r,\xr}}
\newcommand{\card}{\#\!}
\renewcommand{\int}{\textup{int}}			\newcommand{\ext}{\textup{ext}}
\renewcommand{\S}{\mathscr S}						\renewcommand{\L}{\mathscr L}
\newcommand{\K}{\mathscr K}
\renewcommand{\P}{\mathscr P}
\newcommand{\ZZ}{\mathcal Z}
\newcommand{\PS}{\mathscr P \! \mathscr S}			\newcommand{\PL}{\mathscr P \! \mathscr L}
\newcommand{\CP}{\mathscr C \! \mathscr P}
\newcommand{\T}{\mathcal T}

\newcommand{\sumS}{\sideset{}{^*}\sum_{(S_i)_i}}	\newcommand{\sumL}{\sideset{}{^*}\sum_{(L_k)_k}}
\newcommand{\sumSx}{\sideset{}{^*}\sum_{\substack{(S_i)_i \\ \cup_iS_i\ni x}}}
\newcommand{\sumLx}{\sideset{}{^*}\sum_{\substack{(L_k)_k \\ \cup_kL_k\ni x}}}

\newcommand{\sumPLa}{\sideset{}{^*}\sum_{(P_t)_t\in\CP_\La}\!}
\newcommand{\sumPLaE}{\sideset{}{^*}\sum_{\substack{(P_t)_t\in\CP_\La \\ \exists t:\,P_t\in\mathscr E}}}
\newcommand{\sumPLax}{\sideset{}{^*}\sum_{(P_t)_t\in\CP_{\La\setminus x}} \!\!\!\!}
\newcommand{\sumPLay}{\sideset{}{^*}\sum_{(P_t)_t\in\CP_{\La\setminus y}} \!\!\!\!}
\newcommand{\sumPLaxy}{\sideset{}{^*}\sum_{(P_t)_t\in\CP_{\La\setminus x,y}} \!\!\!\!\!\!}

\newcommand{\sumPLadist}{\sideset{}{^*}\sum_{\substack{(P_t)_t\in\CP_\La \\ \exists t:\,\dist_\h(\supp P_t,\,x)\leq1}} \!\!\!\!\!\!\!\!\!\!\!\!}
\newcommand{\sumPLaxdist}{\sideset{}{^*}\sum_{\substack{(P_t)_t\in\CP_{\La\setminus x} \\ \exists t:\,\dist_\h(\supp P_t,\,x)\leq1}} \!\!\!\!\!\!\!\!\!\!\!\!}

\newcommand{\sumPLadistt}{\sideset{}{^*}\sum_{\substack{(P_t)_t\in\CP_\La \\ \exists t:\,\dist_\h(\supp P_t,\,x)\leq1 \\ \exists t':\,\dist_\h(\supp P_{t'},\,y)\leq1}} \!\!\!\!\!\!\!\!\!\!\!\!}
\newcommand{\sumPLaxdistt}{\sideset{}{^*}\sum_{\substack{(P_t)_t\in\CP_{\La\setminus x} \\ \exists t:\,\dist_\h(\supp P_t,\,x)\leq1 \\ \exists t':\,\dist_\h(\supp P_{t'},\,y)\leq1}} \!\!\!\!\!\!\!\!\!\!\!\!}
\newcommand{\sumPLaydistt}{\sideset{}{^*}\sum_{\substack{(P_t)_t\in\CP_{\La\setminus y} \\ \exists t:\,\dist_\h(\supp P_t,\,x)\leq1 \\ \exists t':\,\dist_\h(\supp P_{t'},\,y)\leq1}} \!\!\!\!\!\!\!\!\!\!\!\!}
\newcommand{\sumPLaxydistt}{\sideset{}{^*}\sum_{\substack{(P_t)_t\in\CP_{\La\setminus x,y} \\ \exists t:\,\dist_\h(\supp P_t,\,x)\leq1 \\ \exists t':\,\dist_\h(\supp P_{t'},\,y)\leq1}} \!\!\!\!\!\!\!\!\!\!\!\!}

\newcommand{\nuh}{\nu_\h}							\newcommand{\nuv}{\nu_\v}
\newcommand{\nueven}{\nu_\textup{even}}				\newcommand{\nuodd}{\nu_\textup{odd}}
\newcommand{\Deltao}{\Delta_\textup{orient.}}		\newcommand{\Deltap}{\Delta_\textup{transl.}}	
\newcommand{\fn}{\footnotesize}
\DeclareMathOperator{\dist}{dist}
\DeclareMathOperator{\supp}{supp}

\numberwithin{equation}{section}

\journalname{Preprint}

\begin{document}

\title{A cluster expansion approach to the Heilmann-Lieb liquid crystal model}

\author{Diego Alberici}

\institute{
			University of Bologna - Department of Mathematics \\
            piazza di Porta San Donato 5, Bologna (Italy) \\
            \email{diego.alberici2@unibo.it} }

\date{Received: date / Accepted: date}

\maketitle

\begin{abstract}
A monomer-dimer model with a short-range attractive interaction favoring colinear dimers is considered on the lattice $\Z^2$.
Although our choice of the chemical potentials results in more horizontal than vertical dimers,
the horizontal dimers have no long-range translational order - in agreement with the Heilmann-Lieb conjecture \cite{HLliq}.
\end{abstract}

\section*{Introduction}
A liquid crystal, at low temperatures, displays a long-range order in the orientation of its molecules, while there is no complete ordering in their positions.
In this paper we present a model characterized by these two features.
In particular we consider a monomer-dimer model on the two-dimensional lattice $\Z^2$ characterized by different chemical potentials for horizontal and vertical dimers ($\mu_\h>\mu_\v$ to fix ideas) and by a short-range potential $J>0$ that favors collinear dimers.
We prove that when the parameters satisfy
\eq \label{eq: intro}
\mu_\h>-J \quad\text{and}\quad \mu_\v<-\frac{5}{2}\,J \;,
\eeq
the system has the properties of a liquid crystal.

Onsager \cite{O} was the first to propose hard-rods models in order to explain the existence of liquid crystals.
In 1970 Heilmann and Lieb \cite{HLprl,HLcmp} studied systems of monomer and dimers (hard-rods of length 2) interacting only via the hard-core potential, and proved the absence of phase transitions in great generality.
Then in 1972 they \cite{HLliq} proposed two monomer-dimer models (named \textit{I} and \textit{II}) on the lattice $\Z^2$, where short-range attractive interactions among parallel dimers are considered beyond the hard-core interaction. Heilmann and Lieb claimed that these systems are liquid crystals. In particular they proved the presence of a phase transition, by means of a reflection positivity argument: at low temperature there is orientational order. Moreover they conjectured the absence of complete translational ordering for their models.
A proof of this conjecture for the model \textit{I} was announced in \cite{HLliq} by Heilmann and Kj\ae r, but never appeared.
Letawe, in her thesis \cite{L}, claimed to prove the conjecture by cluster expansion methods, even if the result has never been published in a journal. Letawe's polymers are built starting from contours and the major difficulty seems to arise when she has to deal with a polymer lying in the interior of another one: the two polymers would not be independent. To overcome this problem, ratios of partition functions with different (horizontal or vertical) boundary conditions $Z^\textup v/Z^\textup h$ are introduced, but it is not proved that these ratios are sufficiently small to guarantee the convergence of the cluster expansion.

Numerical simulations related to the Heilmann-Lieb conjecture are performed in \cite{PCF}.
We also mention that, in absence of attractive interaction, systems of sufficiently long hard-rods were proved to display a phase transition and behave like liquid crystals by Disertori and Giuliani \cite{DG}, using a two scales cluster expansion and the Pirogov-Sinai theory.

In the present paper we study a model obtained from the model \textit{I} of Heilmann and Lieb \cite{HLliq}, but while they suppose
\eq \label{eq: intro HL}
\mu_\h=\mu_\v=:\mu \quad\text{and}\quad \mu>-J \;,
\eeq
we assume very different horizontal and vertical potentials as in \eqref{eq: intro}. This choice of the parameters allows us to work with cluster expansion methods, by defining our polymers starting from regions of vertical dimers, instead of contours.
The cluster expansion method permits to rewrite the logarithm of the partition function of a polymer system as a power series of the polymer activities. This expansion entails analyticity results and simplifies considerably the study of the correlation functions, which can be expressed in terms ratios of partition functions.
Clearly the cluster expansion cannot hold in general on the whole space of parameters: it converges only when the polymer activities are small enough to compete with the entropy.
A rigorous study of the conditions of convergence dates back to \cite{GMlMs,GK,R}, by means of Kirkwood-Salsburg type of equations. In this paper we use a criterion proposed by Kotecky and Preiss \cite{KP} in 1986. Afterwards this criterion was compared to the previous ones, was improved and simplified in \cite{BZ,C,D,FP,Ms,U} (for a clear and modern treatment we suggest for example the last work).

The paper is organized as follows. In the section \ref{sec: main} we introduce the model and we state the main results about its liquid crystal properties. In the section \ref{sec: Poly} we show how to rewrite the partition function as a suitable polymer partition function, following in part the ideas of \cite{L}: our polymers turn out to be connected families of \textit{regions} of vertical dimers and \textit{lines} of horizontal dimers and monomers. In the section \ref{sec: CE} we prove that the Kotecky-Preiss condition for the convergence of the cluster expansion is verified when the parameters satisfy \eqref{eq: intro} and the temperature is sufficiently low. Finally in the section \ref{sec: LiqCry} we use the previous sections to prove the results stated in the section \ref{sec: main}.
The appendix \ref{sec: 1D} contains the study of a 1-dimensional monomer-dimer model, that is needed in the section \ref{sec: Poly}.
For the sake of completeness, in the appendix \ref{sec: absCE} we state the general results of cluster expansion needed in the paper.

\section{Definitions and Main Results: the Model and its Liquid Crystal Properties} \label{sec: main}

A \textit{monomer-dimer configuration} on $\Z^2$ can be represented by a bonds\footnote{Two sites $x=(x_\h,x_\v),\,y=(y_\h,y_\v)\in\Z^2$ are \textit{neighbors} ($x\sim y$) if $|x_\h-y_\h|+|x_\v-y_\v|=1\,$. A pair of sites $(x,y)$ is a \textit{bond} if $x,y$ are neighbors. $B(\Z^2)$ denotes the set of bonds.} occupation vector $\alpha\in\{0,1\}^{B(\Z^2)}$ with hard-core interaction, namely:
\eq \label{eq: Poly hardcore}
\sum_{y\sim x} \alpha_{(x,y)} \,\leq\, 1 \quad \forall x\in\Z^2 \;.
\eeq
If $\alpha_{(x,y)}=1$, we say that there is a \textit{dimer} on the bond $(x,y)\,$, or also that there is a dimer at the site $x$; if instead $\alpha_{(x,y)}=0$ for all $y\sim x$, we say that there is a \textit{monomer} on the site $x\,$.
Dimers on $\Z^2$ may have two different orientations: vertical (\textit{v-dimers}) or horizontal (\textit{h-dimers}), according to the orientation of the occupied bond%
\footnote{Two sites $x=(x_\h,x_\v),\,y=(y_\h,y_\v)\in\Z^2$ are \textit{h-neighbors} if $x_\v=y_\v$ and $|x_\h-y_\h|=1$, they are \textit{v-neighbors} if $x_\h=y_\h$ and $|x_\v-y_\v|=1$. A bond $(x,y)\in B(\Z^2)$ is \textit{horizontal} if $x,y$ are h-neighbors, it is \textit{vertical} if $x,y$ are v-neighbors.}.
The model studied in the present paper favors one orientation of the dimers (the horizontal one), both via a chemical potential and via a short-range imitation.

Let $\La$ be a finite sub-lattice of $\Z^2$.
Consider a \textit{horizontal boundary condition}\footnote{The \textit{external boundary} of $\La$ is $\partial^\ext\La:=\{x\in\Z^2\setminus\La\,|\,x\text{ neighbor of }y\in\La\}$. The \textit{internal boundary} of $\La$ is instead $\partial\La\equiv\partial^\int\La:=\{x\in\La\,|\,x\text{ neighbor of }y\in\Z^2\setminus\La\}$. We set $\bar\La:=\La\cup\partial^\ext\La\,$.}, namely we assume that every site of $\Z^2\setminus\Lambda$ has a h-dimers (with either free or fixed positions).
Denote by $\D_\La^\h$ the set of monomer-dimer configurations on $\La$ (we allow also dimers toward the exterior\footnote{Namely we allow dimers having one endpoint in $\La$ and one in $\Z^2\setminus\La\,$.
}) which are compatible with the selected horizontal boundary condition.

The Hamiltonian, or energy, of a monomer-dimer configuration is defined as
\eq \label{eq: Poly hamilt} \begin{split} 
H_\La :=\; & \frac{\muh+J}{2}\,\card\left\{\parbox{6em}{\fn sites of $\La$ with monomer}\right\} \,+\,
\frac{\muh-\muv}{2}\,\card\left\{\parbox{6em}{\fn sites of $\La$ with v-dimer}\right\} \,+\\[4pt]
&+ \frac{J}{2}\left( \card\left\{\parbox{10em}{\fn sites of $\bar\La$ with h-dimer but h-neighbor also to a v-dimer or a monomer}\right\} + \card\left\{\parbox{10em}{\fn sites of $\bar\La$ with v-dimer but v-neighbor also to a h-dimer or a monomer}\right\} \right).
\end{split}\eeq
We assume that the parameters appearing in the Hamiltonian satisfy
\eq \muh>-J\,,\quad \muh\geq\muv\,,\quad J>0 \;.\eeq
In this way, if the horizontal boundary condition with free positions is chosen%
\footnote{Also fixed positions work, provided that the positions of the two h-dimers at the endpoints of each horizontal line of $\La$ allow a pure dimer configuration on that line.}%
, then the \textit{ground states} in $\D_\La^\h$ (i.e. the configurations minimizing the energy under the given condition) are exactly the configurations where every site has a h-dimer.
The partition function of the system is
\eq  \label{eq: Poly Z}
Z_\La^\h \,:=\, \sum_{\alpha\in\D_\La^\h} e^{-\beta H_\La(\alpha)} \eeq
where the parameter $\beta>0$ is the inverse temperature.

\begin{remark}
We want to show that the Hamiltonian \eqref{eq: Poly hamilt} essentially corresponds to the model \textit{I} introduced by Heilmann and Lieb in \cite{HLliq}, except for the important fact that we allow the horizontal and vertical dimer potentials $\muh,\,\muv$ to be different, while they take $\muh=\muv=\mu\,$.
We can introduce another Hamiltonian (that maybe is written in a more natural way; see fig.\ref{fig1}):
\eq \label{eq: Poly hamilt original} \begin{split}
\widetilde H_\La \,:=\,& -\,\muh\,\card\left\{\parbox{5.6em}{\fn h-dimers in $\La$}\right\} \,
 -\,\muv\,\card\left\{\parbox{5.6em}{\fn v-dimers in $\La$}\right\} \;+\\
 & -\,J\,\card\left\{\parbox{8.2em}{\fn pairs of neighboring colinear dimers in $\La$}\right\}
\end{split} \eeq

\begin{figure}[h] 
\centering
\includegraphics[scale=0.82]{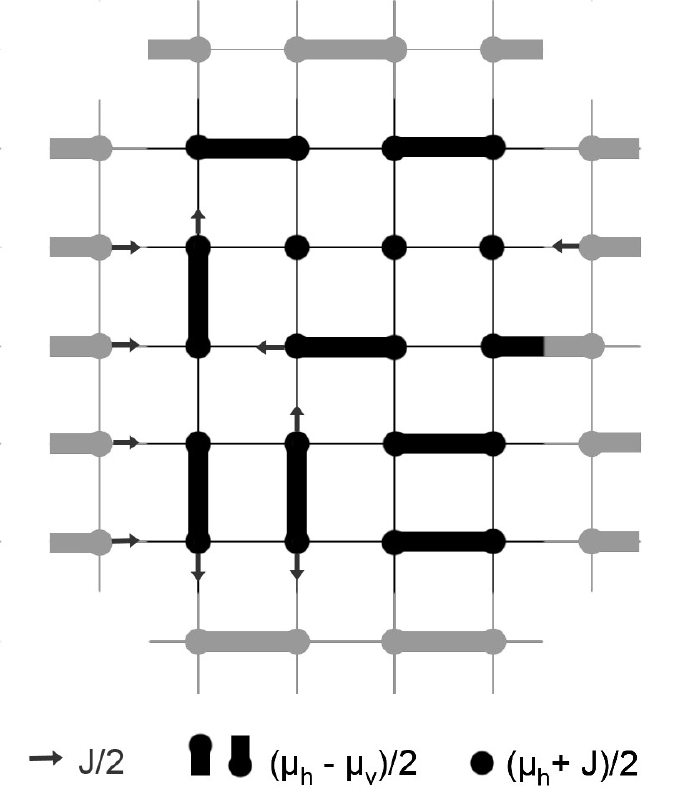}
\includegraphics[scale=0.82]{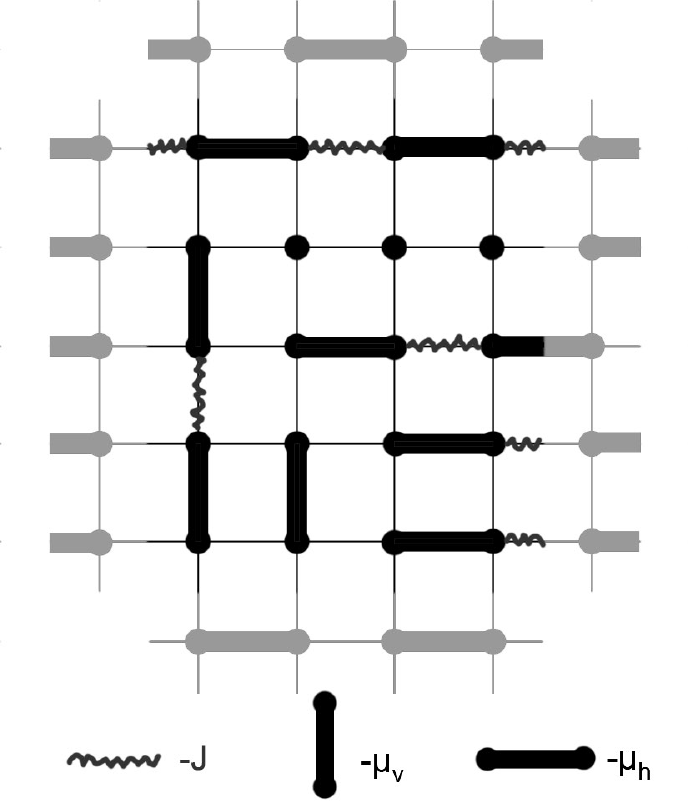}
\caption{The same monomer-dimer configuration on the lattice $\La$ and the corresponding energies in accordance to the Hamiltonian \eqref{eq: Poly hamilt} \textup{(on the left)} and to the Hamiltonian \eqref{eq: Poly hamilt original} \textup{(on the right)}. A horizontal boundary condition is drawn in grey.}
\label{fig1}
\end{figure}
\noindent The monomer-dimer model \textit{I} in \cite{HLliq} is given by the Hamiltonian \eqref{eq: Poly hamilt original} with $\muh=\muv=\mu$, when $\La$ is a rectangular lattice of even sides lengths with periodic boundary conditions (torus).
It is easy to show that when $\La$ is a torus the two Hamiltonians \eqref{eq: Poly hamilt}, \eqref{eq: Poly hamilt original} describe the same model; indeed they only differ by an additive constant which does not affect the Gibbs measure:
\eq
\widetilde H_\La \,+\, \frac{\muh+J}{2}\,|\La| \;=\; H_\La
\eeq
since
\[\begin{split} & |\La|-2\,\card\left\{\parbox{5.6em}{\fn h-dimers in $\La$}\right\} \,=\,
|\La|-\card\left\{\parbox{6em}{\fn sites in $\La$ with h-dimer}\right\} \,=\\
&=\, \card\left\{\parbox{6em}{\fn sites in $\La$ with monomer}\right\} + \card\left\{\parbox{6em}{\fn sites in $\La$ with v-dimer}\right\} \;;\end{split}\]
\[ 2\,\card\left\{\parbox{5.6em}{\fn v-dimers in $\La$}\right\} \,=\, \card\left\{\parbox{6em}{\fn sites in $\La$ with v-dimer}\right\} \;;\]
\[\begin{split} & |\La|-2\,\card\left\{\parbox{8.2em}{\fn pairs of neighboring colinear dimers in $\La$}\right\} =\,
|\La|-\card\left\{\parbox{13em}{\fn sites in $\La$ with h-dimer (v-dimer) and h-neighbor (v-neighbor) to another h-dimer (v-dimer)}\right\} =\\
&=\, \card\left\{\parbox{6em}{\fn sites in $\La$ with monomer}\right\} + \card\left\{\parbox{13em}{\fn sites in $\La$ with h-dimer (v-dimer) and h-neighbor (v-neighbor) also to something different}\right\} \;.\end{split}\]
On the other hand when $\La$ has horizontal boundary conditions the two Hamiltonians \eqref{eq: Poly hamilt}, \eqref{eq: Poly hamilt original} are not exactly equivalent. Indeed it holds%
\footnote{$\partial_\v,\,\partial_\h$ denote respectively the \textit{vertical}, \textit{horizontal component} of the boundary; e.g. $\partial_\v\La:=\{x\in\La\,|\,x\text{ h-neighbor of }y\in\Z^2\setminus\La\}$ and $\partial_\h\La:=\{x\in\La\,|\,x\text{ v-neighbor of }y\in\Z^2\setminus\La\}$.}%
\eq
\widetilde H_\La \,+\, \frac{\muh+J}{2}\,|\La| \,+\, \frac{J}{2}\;\card\left\{\parbox{7em}{\fn sites in $\partial_\v^\int\La$ without h-dimer}\right\} \,=\, H_\La
\eeq
when the following conventions are adopted in the definition \eqref{eq: Poly hamilt original}: if only half a dimer is in $\La$ while the other half is in $\Z^2\setminus\La$, it counts $\frac{1}{2}$; if only one dimer of a pair of neighboring colinear dimers is in $\La$, while the other one is in $\Z^2\setminus\La$, this pair counts $\frac{1}{2}$.
\end{remark}

The monomer-dimer model that we have introduced, in a certain region of the parameters corresponding to large horizontal potential, small vertical potential and low temperature, behaves like a \textit{liquid crystal}.
This means that the model exhibits an order in the orientation of the molecules (dimers), while there is no complete order in their positions.

The following results will give a precise mathematical meaning to these statements. First we introduce some observables attached to the sites, asking questions as \virg{Is there a horizontal dimer at site $x$?}, \virg{If so, is it positioned to the left or to the right of $x$?}. To measure the absence or presence of some kind of order, at a microscopic level we study the expectations and the covariances of these quantities according to the Gibbs measure, while at a macroscopic level we introduce a suitable \textit{order parameter} and study its expectation and possibly its variance\footnote{When the expectation of the order parameter is zero but the variance is not, a small perturbation can lead to a spontaneous order of the system.}.

Define the following local observables\footnote{We say that the site $x$ has a \textit{left-dimer} if there is a dimer on the bond $\big(x,x-(1,0)\big)\,$, a \textit{right-dimer} if there is a dimer on the bond $\big(x,x+(1,0)\big)\,$.}
\begin{gather}
\label{eq: LiqCry deffhfv}
f_{\h,x} \,:=\, \1\big(x\text{ has a h-dimer}\big) \;, \quad
f_{\v,x} \,:=\, \1\big(x\text{ has a v-dimer}\big) \;; \\[2pt]
\label{eq: LiqCry defflfr}
f_{\l,x} \,:=\, \1\big(x\text{ has a left-dimer}\big)\;, \quad
f_{\r,x} \,:=\, \1\big(x\text{ has a right-dimer}\big)\;.
\end{gather}
Clearly $f_{\h,x}=f_{\l,x}+f_{\r,x}$ and $f_{\h,x}+f_{\v,x}\leq 1\,$.
In the following we denote the Gibbs expectation of any observable $f$ by
\[ \langle f \rangle_\La^\h \,:=\, \frac{1}{Z_\La^\h}\, \sum_{\al\in\D_\La^\h} f(\al)\, e^{-\beta H_\La(\al)} \;.\]

We denote by $N$ the minimal distance between any two vertical components of the boundary of $\La$ and our only assumption on the shape of $\La$ is that $N\to\infty$ as $\La\nearrow\Z^2\,$.
To fix ideas one could think that $\La$ is a rectangle (in this case $N$ would be simply its horizontal side length), but actually we will need to consider also non-simply connected regions.

There exists $\beta_0>0$ depending on $\muh,\,\muv,\,J$ only and $N_0(\beta)$ depending on $\beta,\,\muh,\,J$ only such that the following results hold true.

\begin{theorem}[Microscopic expectations] \label{thm: LiqCry orientorder}
Assume that $J>0$, $\muh+J>0$ and $2\muv+5J<0$. Let $\beta>\beta_0$. Let $\La\subset\Z^2$ finite having $N>N_0(\beta)$.
Let $x\in\La$ such that $\dist_\h(x,\partial\La)>N_0(\beta)$.
Then
\eq \label{eq: LiqCry flrbound}
\langle f_{\l,x} \rangle_\La^\h \,\geq\, \frac{1}{2}-e^{-\beta\frac{\muh+J}{2}} \quad,\quad
\langle f_{\r,x} \rangle_\La^\h \,\geq\, \frac{1}{2}-e^{-\beta\frac{\muh+J}{2}} \;.
\eeq
As a consequence:
\begin{gather}
\label{eq: LiqCry fhbound}
\langle f_{\h,x} \rangle_\La^\h \,\geq\, 1-2\,e^{-\beta\frac{\muh+J}{2}} \;;\\[2pt]
\label{eq: LiqCry fl-frbound}
\big|\, \langle f_{\r,x} \rangle_\La^\h - \langle f_{\l,x} \rangle_\La^\h \,\big| \,\leq\, 2\,e^{-\beta\frac{\muh+J}{2}} \;.
\end{gather}
\end{theorem}

\begin{theorem}[Microscopic covariances] \label{thm: LiqCry corrdecay}
Assume that $J>0$, $\muh+J>0$ and $2\muv+5J<0$. Let $\beta>\beta_0$. Let $\La\subset\Z^2$ finite such that $N>N_0(\beta)$.
Let $x,y\in\La$ such that $\dist_\h(x,\partial\La)>N_0(\beta)$, $\dist_\h(y,\partial\La)>N_0(\beta)$ and $\dist_\h(x,y)>N_0(\beta)$.
Then:
\begin{gather}
\label{eq: LiqCry flfl} \big| \langle f_{\l,x}\,f_{\l,y} \rangle_\La^\h - \langle f_{\l,x} \rangle_\La^\h\, \langle f_{\l,y} \rangle_\La^\h \big| \,\leq\, \frac{9m}{16}\; e^{-\frac{m}{4}(\dist_{\Z^2}(x,y)-1)} \;,\\[4pt]
\label{eq: LiqCry frfr} \big| \langle f_{\r,x}\,f_{\r,y} \rangle_\La^\h - \langle f_{\r,x} \rangle_\La^\h\, \langle f_{\r,y} \rangle_\La^\h \big| \,\leq\, \frac{9m}{16}\; e^{-\frac{m}{4}(\dist_{\Z^2}(x,y)-1)} \;,\\[4pt]
\label{eq: LiqCry flfr} \big| \langle f_{\l,x}\,f_{\r,y} \rangle_\La^\h - \langle f_{\l,x} \rangle_\La^\h\, \langle f_{\r,y} \rangle_\La^\h \big| \,\leq\, \frac{9m}{16}\; e^{-\frac{m}{4}(\dist_{\Z^2}(x,y)-1)} \;.
\end{gather}
The definition of $m$ is clarified in the Appendix (lemma \ref{lem: 1D eigenval ratio}); anyway it can be sufficient to know that $m = e^{-\beta\frac{\muh+3J}{2}}\,(1+o(1))$ as $\beta\to\infty$.
\end{theorem}

The density of lattice sites occupied by h-dimers/v-dimers is respectively:
\eq \label{eq: LiqCry defnuhnuv}
\nuh \,:=\, \frac{1}{|\La|}\,\sum_{x\in\La}\, f_{\h,x} \quad,\quad
\nuv \,:=\, \frac{1}{|\La|}\,\sum_{x\in\La}\, f_{\v,x} \;.
\eeq
A parameter measuring the orientational order of the dimers is
\eq \label{eq: LiqCry defDeltaorient}
\Deltao \,:=\, \nuh-\nuv \;.
\eeq

\begin{corollary}[Orientational Order Parameter] \label{cor: LiqCry orientorder}
Assume that $J>0$, $\muh+J>0$ and $2\muv+5J<0$. Let $\beta>\beta_0$. Let $\La\subset\Z^2$ finite, having $N>2\,N_0(\beta)\,$.
Then
\eq \label{eq: LiqCry horizdens}
\langle \Deltao \rangle_{\La}^{\h} \,\geq\, \Big(1-2\,\frac{N_0(\beta)}{N}\Big)\, \big(1-4\,e^{-\beta\frac{\muh+J}{2}}\big) \;.
\eeq
Hence
\eq \label{eq: LiqCry horizdens limit}
\lim_{\beta\nearrow\infty}\liminf_{\La\nearrow\Z^2}\; \langle \Deltao \rangle_{\La}^{\h} \,=\, 1 \;.
\eeq
\end{corollary}

The corollary \ref{cor: LiqCry orientorder} shows that fixing $\beta$ sufficiently large and then choosing $\La$ sufficiently big (more precisely the distance $N$ between vertical components of $\partial\La$ must be large enough), the average density of sites occupied by h-dimers is arbitrarily close to $1\,$: in other terms the system is oriented along the horizontal direction.

The majority of sites is occupied by h-dimers. But there can still be some freedom, indeed we may distinguish the h-dimers in two classes according to their positions: a \textit{h-dimer} is called \textit{even} (resp. \textit{odd}) if its left endpoint has even (resp. odd) horizontal coordinate.
The density of lattice sites occupied by even/odd h-dimers is respectively:
\eq \label{eq: LiqCry defnuevennuodd} \begin{split}
\nueven \,&:=\, \frac{1}{|\La|}\,\sum_{x\in\La}\, \1\big(x\text{ has an even h-dimer}\big) \,=\,
\frac{2}{|\La|} \sum_{x\in\La\atop x_\h\!\text{ even}} f_{\r,x} \;,\\
\nuodd \,&:=\, \frac{1}{|\La|}\,\sum_{x\in\La}\, \1\big(x\text{ has an odd h-dimer}\big) \,=\,
\frac{2}{|\La|} \sum_{x\in\La\atop x_\h\!\text{ even}} f_{\l,x}  \;.
\end{split} \eeq
A parameter measuring the translational order of the h-dimers is
\eq \label{eq: LiqCry defDelta}
\Deltap:=\nueven-\nuodd \;.
\eeq

\begin{corollary}[Translational Order Parameter. Part I] \label{cor: LiqCry posorder1}
Assume that $J>0$, $\muh+J>0$ and $2\muv+5J<0$. Let $\beta>\beta_0$. Let $\La\subset\Z^2$ finite such that $N>2\,N_0(\beta)\,$.
Then
\eq \label{eq: LiqCry evenodd average}
\big| \langle \Deltap \rangle_{\La}^{\h} \big| \,\leq\,
\Big(1-2\frac{N_0(\beta)}{N}\Big)\, 2\,e^{-\beta\frac{\muh+J}{2}} + 2\,\frac{N_0(\beta)}{N} 
\eeq
Hence
\eq \label{eq: LiqCry evenodd averagelimit}
\lim_{\beta\nearrow\infty}\limsup_{\La\nearrow\Z^2}\; \big|\langle \Deltap \rangle_{\La}^{\h}\big| \,=\, 0 \;.
\eeq
\end{corollary}

\begin{corollary}[Translational Order Parameter. Part II] \label{cor: LiqCry posorder2}
Assume that $J>0$, $\muh+J>0$ and $2\muv+5J<0$. Let $\beta>\beta_0$. Let $\La\subset\Z^2$ finite such that $N>2\,N_0(\beta)\,$.
Then
\eq \label{eq: LiqCry evenodd variance}
\big\langle (\Deltap)^2 \big\rangle_{\La}^{\h} \,-\, \big(\langle \Deltap \rangle_{\La}^{\h}\big)^2 \,\leq\,
\frac{1}{|\La|}\,\frac{9m}{(1-e^{-\frac{m}{4}})^2} \,+\, \frac{N_0(\beta)}{N}\,\Big(6-8\frac{N_0(\beta)}{N}\Big) \;.
\eeq
Hence for fixed $\beta>\beta_0$
\eq \label{eq: LiqCry evenodd variancelimit}
\lim_{\La\nearrow\Z^2} \big\langle (\Deltap)^2 \big\rangle_{\La}^{\h} \,-\, \big(\langle \Deltap \rangle_{\La}^{\h}\big)^2  \,=\, 0\;.
\eeq
\end{corollary}

The corollaries \ref{cor: LiqCry posorder1}, \ref{cor: LiqCry posorder2} 
show that fixing $\beta$ sufficiently large and then choosing $\La$ sufficiently big (in particular the distance between different components of $\partial_\v\La$ must be big enough), the mean value and the variance of the difference between the density of even h-dimers and the density of odd h-dimers are arbitrarily close to zero. In other terms, at large but finite $\beta$, there is not a spontaneous translational order for the h-dimers. 

\begin{remark} \label{rk: LiqCry posorder}
The bounds \eqref{eq: LiqCry evenodd average} hold for any kind of horizontal boundary conditions, but in some particular cases it is possible to obtain a better result by a symmetry argument.
Assume that $\La$ is a rectangle with $N+1$ sites in each horizontal side.
If $N+1$ is odd, by choosing \textit{horizontal dimers with free positions at the boundary} one obtains 
\eq
\langle \Deltap \rangle_{\La}^{\h} \,=\, \langle \nueven \rangle_{\La}^{\h} - \langle \nuodd \rangle_{\La}^{\h} \,=\, 0
\eeq
for all parameters $\beta,\,J,\,\muh,\,\muv\,$.
To prove it consider the reflection on $\La$ with respect to the vertical axis at distance $\frac{N}{2}$ from $\partial_\v\La$: this transformation induces a bijection $T\!:\D_\La^\h\rightarrow\D_\La^\h\,$. It is easy to check that 
$H_\La(T(\al))=H_\La(\al)\,$, $\nueven(T(\al))=\nuodd(\al)\,$, $\nuodd(T(\al))=\nueven(\al)$ for all $\al\in\D_\La^\h\,$.

\noindent On the other hand if $N+1$ is even, by choosing \textit{periodic boundary conditions} one still obtains
\eq
\langle \Deltap \rangle_{\La}^{\textup{per.}} \,=\, 0
\eeq
for all parameters $\beta,\,J,\,\muh,\,\muv\,$.
To prove it one can consider the reflection on $\La$ with respect to two vertical axis at distance $\frac{N+1}{2}$ from each other: it induces a bijection from $\D_\La^\textup{per.}$ to itself having all the previous properties.
\end{remark}

\section{Polymer Representation} \label{sec: Poly}
In this section we show how to rewrite the partition function $Z_\La^\h$ as a polymer partition function of type \eqref{eq: absCE Z}. This representation will be suitable for applying the cluster expansion machinery (see Appendix \ref{sec: absCE}) in a regime of large horizontal potential, small vertical potential and low temperature.

\begin{figure}[h] 
\centering
\includegraphics[scale=0.75]{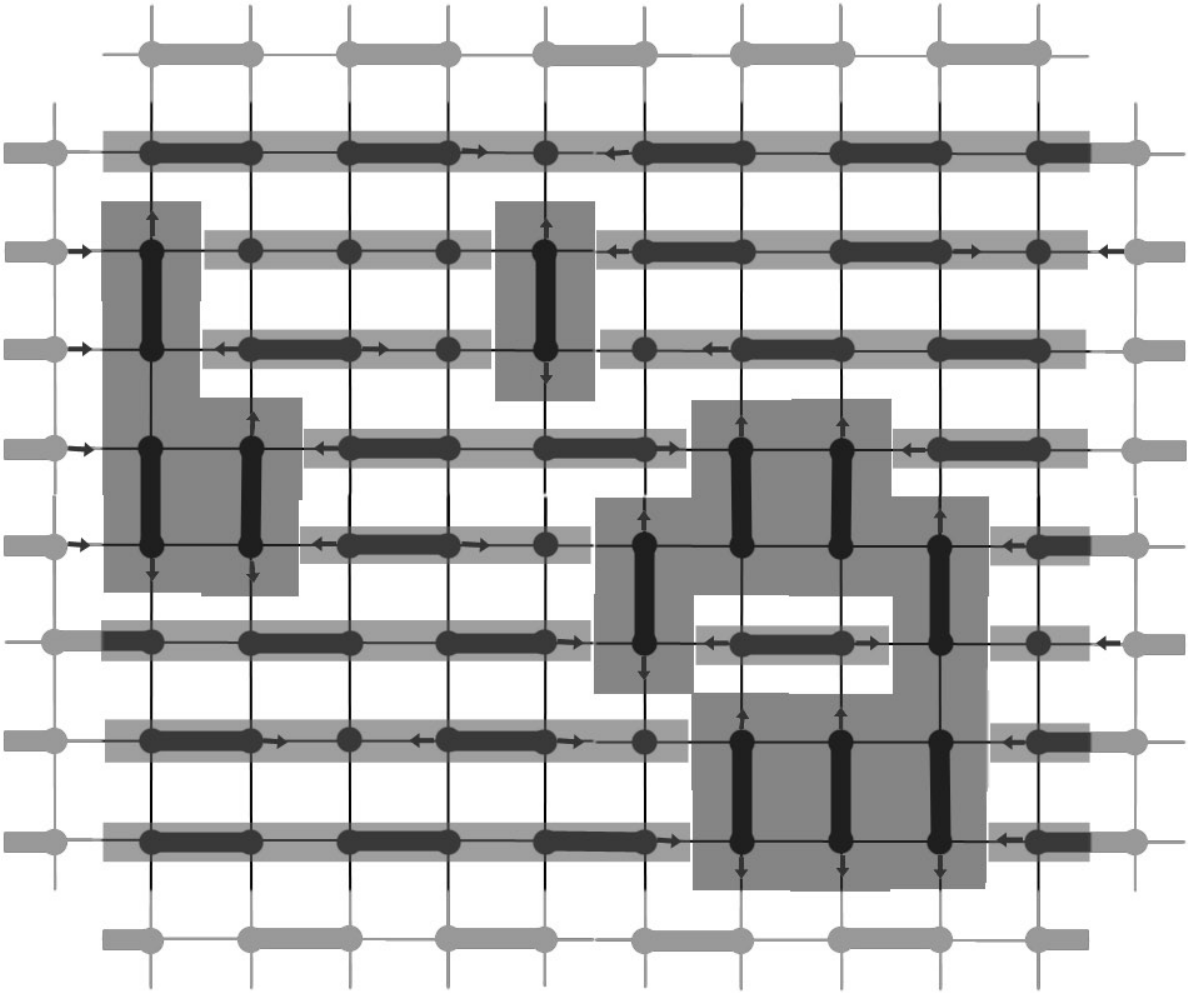}
\caption{A monomer-dimer configuration on $\La$ and the corresponding \textup{regions} $S_1,S_2,S_3$ and \textup{lines} $L_1,\dots,L_{15}\in\L_\La(\cup_iS_i)$. Given the positions of the regions, the configurations on the lines are mutually independent: the \textup{arrows} represent the energy contributions of type $J/2$. A horizontal boundary condition is drawn.}
\label{fig2}
\end{figure}

We start by isolating the \virg{few} vertical dimers. Associate to each monomer-dimer configuration $\al\in\D_\La^\h$ the set
\[ V=V(\al):=\{x\in\La \;|\; x\text{ has a v-dimer according to }\al\} \;.\]
Partition $V$ into its connected components (as a sub-graph of the lattice\footnote{On any graph the distance between two objects is defined as the length of the shortest path connecting them. In particular $\dist_{\Z^2}(S,S'):=\inf_{x\in S,\,y\in S'}\dist_{\Z^2}(x,y)$ for all $S,S'\subset\Z^2$ and $\dist_{\Z^2}(x,y):=|x_\h-y_\h|+|x_\v-y_\v|$ for all $x=(x_\h,x_\v),\,y=(y_\h,y_\v)\in\Z^2\,$.} $\Z^2$):
\[ V = \bigcup_{i=1}^n S_i \quad,\quad S_i\in\S_\La\ \forall\,i\quad,\quad \dist_{\Z^2}(S_i,S_j)>1\ \forall\,i\neq j \]
where the family $\S_\La$ is defined by
\eq \label{eq: Poly defS} \begin{split}
S\in\S_\La \ \overset{\text{def}}\Leftrightarrow\ \,&
S\subseteq\La \;,\; S\neq\emptyset \;,\; S\text{ connected (as a sub-graph of $\Z^2$)} \;,\\
& \text{every maximal vertical segment of $S$ has an even number}\\&\text{of sites}\;,\\
& \text{$S$ does not contains those sites of $\partial_\v^\int\La$ that necessarily}\\&\text{have a h-dimer because of the boundary conditions.} \end{split}\eeq
The knowledge of the set $V$ (or equivalently of $S_1,\dots,S_n$) does not determine completely the configuration $\al$ of the system, since on $\La\setminus V$ there can be both h-dimers and monomers. Anyway a fundamental feature of the model is that the system on $\La\setminus V$ can be partitioned into independent 1-dimensional systems.
Introduce the family $\L_\La(V)$ defined by
\eq \label{eq: Poly defL}
L\in\L_\La(V) \ \overset{\text{def}}\Leftrightarrow\ L\text{ is a maximal horizontal line of } \La\setminus V \,. \eeq
The Hamiltonian \eqref{eq: Poly hamilt} rewrites as
\[\begin{split} H_\La \,=\,&
\sum_{i=1}^n\left(\frac{\muh-\muv}{2}\,|S_i| \,+\, \frac{J}{2}\,|\partial_\h S_i| \,+\, \frac{J}{2}\,|\partial_\v S_i\cap\partial\La| \right) \;+\\
&\!\!\!\!\!\!+\!\!\!\!\!\!\! \sum_{\;L\in\L_\La(\cup_iS_i)}\!\!\! \left( \frac{\muh+J}{2}\,\card\left\{\parbox{6em}{\fn sites of $L$ with monomer}\right\} \,+\, \frac{J}{2}\,\card\left\{\parbox{9.5em}{\fn sites of $L$ with h-dimer but h-neighbor also to a monomer or to $\cup_iS_i$}\right\} \right).  \end{split}\]
%
Hence the partition function \eqref{eq: Poly Z} rewrites as (see fig.\ref{fig2})
\eq \label{eq: Poly Z Si}
Z_\La^\h \,=\,
\sum_{n\geq0}\, \frac{1}{n!}\!\!\!\! \sum_{\substack{S_1,\dots,S_n\in\S_\La\\ \,\dist(S_i,S_j)>1\,\forall i\neq j}} \prod_{i=1}^n\, e^{-\beta\left(\frac{\muh-\muv}{2}|S_i|\,+\,\frac{J}{2}|\partial_\h S_i|\,+\,\frac{J}{2}|\partial_\v S_i\cap\partial\La|\right)}\!\!\!\! \prod_{\,L\in\L_\La(\cup_iS_i)}\!\!\!\!\!\! Z_L \eeq
where $Z_L$ is the monomer-dimer partition function of the line $L$, considered as a sub-lattice of the 1-dimensional lattice $\Z$, with suitable boundary conditions:
\eq \label{eq: Poly Zline}
Z_L \,:=\, \sum_{\al_L\in\D_L} e^{-\beta H_L(\al_L)}\ e^{\Ixl(\alpha_{\xl})}\ e^{\Ixr(\alpha_{\xr})} \;.\eeq

\noindent An explanation of the notations introduced in \eqref{eq: Poly Zline} is required.
$\D_L$ denotes the set of monomer-dimer configurations on $L$ (dimers can only be horizontal, external dimers at the endpoints of $L$ are allowed); 
\[ H_L \,:=\, \frac{\muh+J}{2}\,\card\left\{\parbox{6em}{\fn sites of $L$ with monomer}\right\} \,+\, \frac{J}{2}\,\card\left\{\parbox{9em}{\fn sites of $L$ with dimer but h-neighbor also to a monomer}\right\} \;;\]
$\xl,\,\xr$ denote respectively the left, right endpoint of the line $L$ (which eventually may coincide): observe%
\footnote{$\partial_\l,\,\partial_\r$ denote respectively the \textit{left}, \textit{right component} of the vertical boundary; e.g. $\partial_\l\La:=\{x\in\La\,|\,x-(1,0)\in\Z^2\setminus\La\}$ and $\partial_\r\La:=\{x\in\La\,|\,x+(1,0)\in\Z^2\setminus\La\}$.}
that because of \eqref{eq: Poly defL}
\begin{gather} \label{eq: Poly endptl}
\bigcup_{L\in\L_\La(\cup_iS_i)}\!\!\xl(L) \;=\; \bigg( \big(\cup_i\partial_\r^\ext S_i\big)\cap\La \bigg)\; \sqcup\; \bigg( \partial_\l\La \setminus\cup_i\,\partial_\l S_i \bigg) \;,\\
\label{eq: Poly endptr}
\bigcup_{L\in\L_\La(\cup_iS_i)}\!\!\xr(L) \;=\; \bigg( \big(\cup_i\partial_\l^\ext S_i\big)\cap\La \bigg)\; \sqcup\; \bigg( \partial_\r\La \setminus\cup_i\,\partial_\r S_i \bigg) \;; 
\end{gather}
finally%
\footnote{The possible states of a site $x\in L$ are three: ``l''=\textit{left-dimer} namely a dimer on the bond $\big(x,x-(1,0)\big)\,$,  ``r''=\textit{right-dimer} namely a dimer on the bond $\big(x,x+(1,0)\big)\,$, ``m''=\textit{monomer}. Here we think $\Ixl,\,\Ixr$ as vectors: $\Ixl=\bmat \Ixl(l) & \Ixl(r) & \Ixl(m) \emat$ and $\Ixr=\bmat \Ixr(l) & \Ixr(r) & \Ixr(m) \emat\,$.}
\eq \label{eq: Poly BCendptl} \begin{split}
&\text{if }\xl\in\cup_i\,\partial_\r^\ext S_i \ \Rightarrow\ \Ixl:= \bmat -\infty & -\beta\frac{J}{2} & 0 \emat \\
&\text{if }\xl\in\partial_\l\La\,, \text{ on $\xl\!-\!(1,0)$ it is fixed a l-dimer} \ \Rightarrow\ \Ixl:= \bmat -\infty & 0 & -\beta\frac{J}{2} \emat  \\
&\text{if }\xl\in\partial_\l\La\,, \text{ on $\xl\!-\!(1,0)$ it is fixed a r-dimer} \ \Rightarrow\ \Ixl:= \bmat 0 & -\infty & -\infty \emat \\
&\text{if }\xl\in\partial_\l\La\,, \text{ on $\xl\!-\!(1,0)$ there is a free h-dimer} \ \Rightarrow\ \Ixl:= \bmat 0 & 0 & -\beta\frac{J}{2} \emat
\end{split} \eeq
and, similarly,
\eq \label{eq: Poly BCendptr} \begin{split}
&\text{if }\xr\in\cup_i\,\partial_\l^\ext S_i \ \Rightarrow\ \Ixr:= \bmat -\beta\frac{J}{2} & -\infty & 0 \emat \\
&\text{if }\xr\in\partial_\r\La\,, \text{ on $\xr\!+\!(1,0)$ it is fixed a r-dimer} \ \Rightarrow\ \Ixr:= \bmat 0 & -\infty & -\beta\frac{J}{2} \emat  \\
&\text{if }\xr\in\partial_\r\La\,, \text{ on $\xr\!+\!(1,0)$ it is fixed a l-dimer} \ \Rightarrow\ \Ixr:= \bmat -\infty & 0 & -\infty \emat \\
&\text{if }\xr\in\partial_\r\La\,, \text{ on $\xr\!+\!(1,0)$ there is a free h-dimer} \ \Rightarrow\ \Ixr:= \bmat 0 & 0 & -\beta\frac{J}{2} \emat .
\end{split} \eeq
The 1-dimensional systems described by $Z_L$, $L\in\L_\La(\cup_iS_i)\,$, are studied in the Appendix \ref{sec: 1D}.

\begin{figure}[h]
\centering
\includegraphics[scale=0.5]{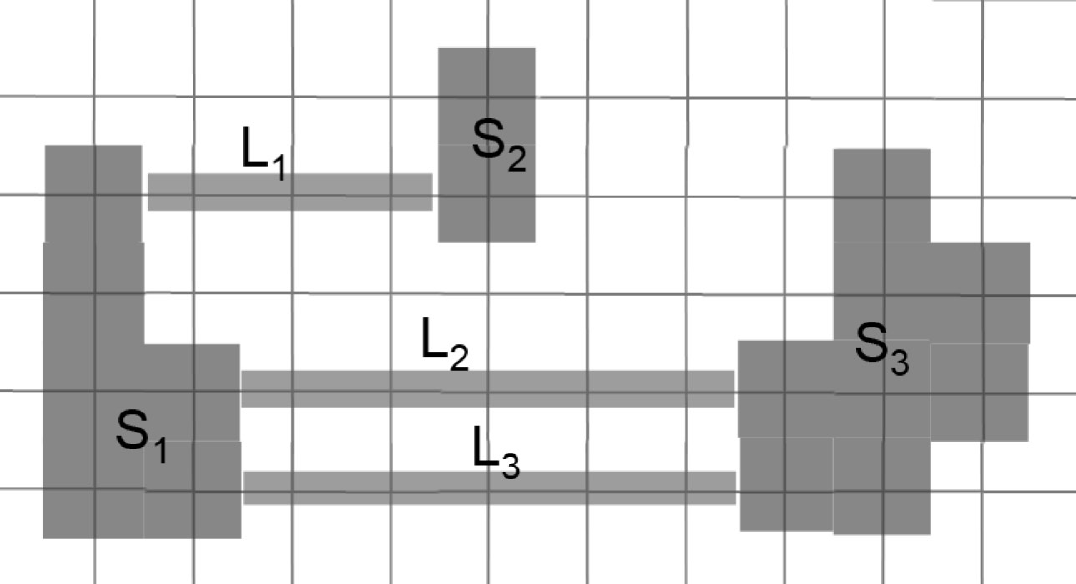}\quad
\includegraphics[scale=0.579]{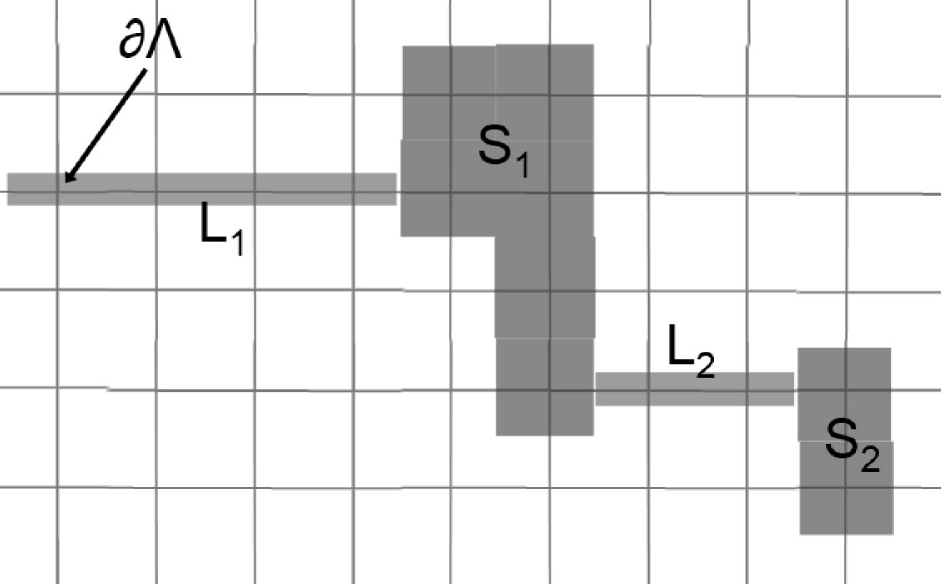}\\[10pt]
\includegraphics[scale=0.5]{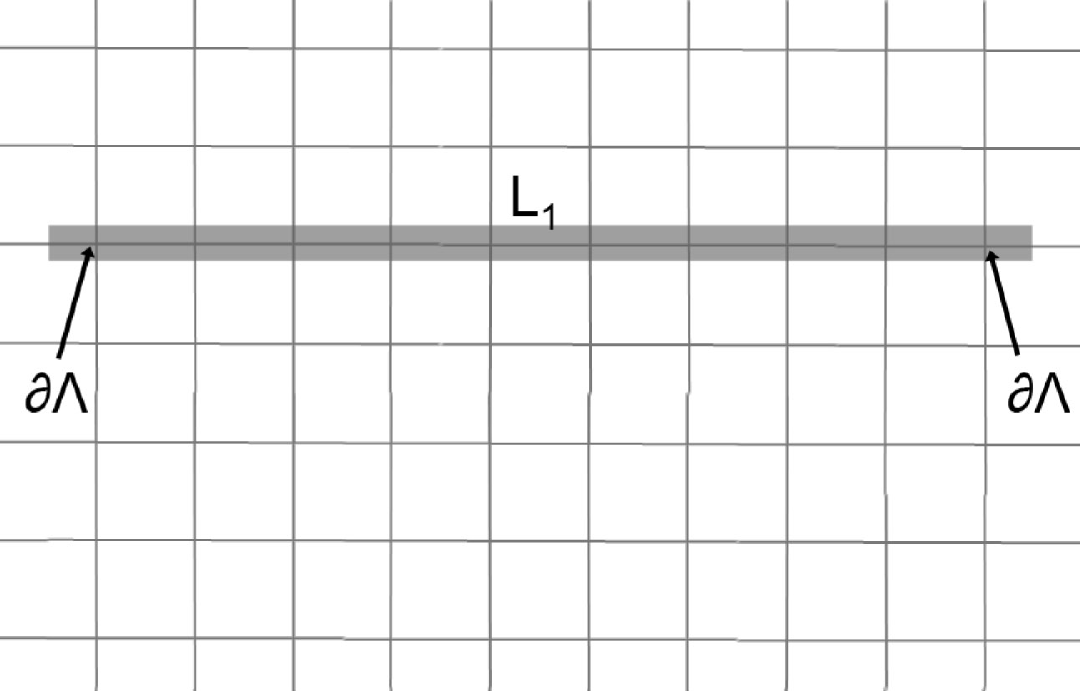}\quad 
\includegraphics[scale=0.611]{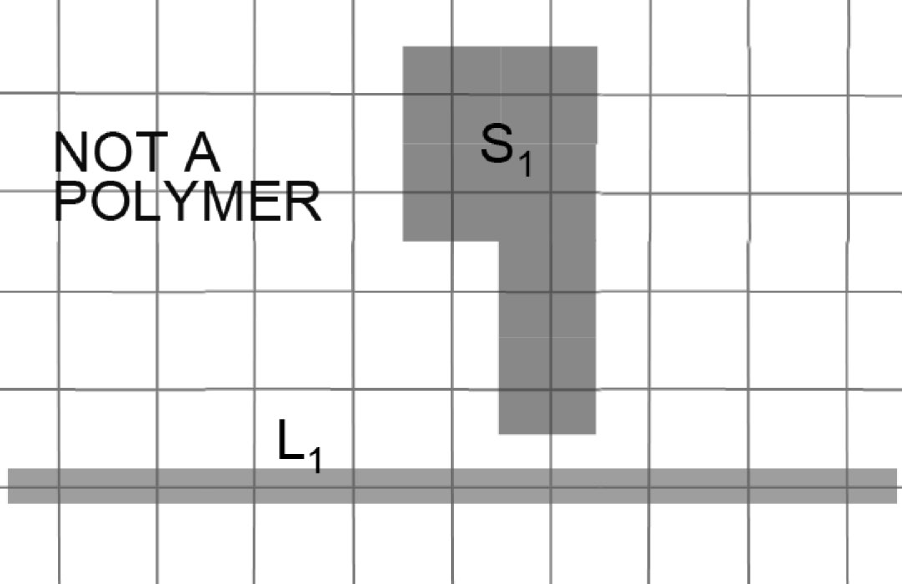}
\caption{The first three pictures represent three different examples of polymers $P\in\P_\La$. The set represented in the last picture is not a unique polymer since it is not connected in $\widetilde\Z^2$ (even if it is connected in $\Z^2$).}
\label{fig3}
\end{figure}

In the form \eqref{eq: Poly Z Si} of $Z_\La^\h$, the weight of the regions $(S_1,\dots,S_n)$ is not a product of the weights of each region $S_i$, because of the lines $L$ connecting different regions. Therefore the regions $S_i\in\S_\La$ are not a good choice for a polymer representation of the model.
In order to decouple some regions from some other ones, it is possible to do a simple trick.
It is convenient to deal in different ways with the endpoints lying on $\partial^\ext S_i$ and those on $\partial\La$; hence given a line $L\in\L_\La(\cup_i S_i)$ we set
\begin{gather*}
\epsxl := \1\left(\xl\in(\cup_i\,\partial_\r^\ext S_i)\cap\La\right) \;,\quad
\etaxl := 1\!-\epsxl \overset{\eqref{eq: Poly endptl}}{=} \1\left(\xl\in(\partial_\l\La)\setminus\cup_i\,\partial_\l S_i\right) \;;\\
\epsxr \!:= \1\left(\xr\in(\cup_i\,\partial_\l^\ext S_i)\cap\La\right) \,,\quad
\etaxr \!:= 1\!-\epsxr \!\overset{\eqref{eq: Poly endptr}}{=} \1\left(\xr\in(\partial_\r\La)\setminus\cup_i\,\partial_\r S_i\right) \,.
\end{gather*}
Using the notations of the Appendix \ref{sec: 1D}, given a line $L\in\L_\La(\cup_i S_i)$ we introduce the two vectors representing the boundary conditions outside its endpoints $\xl,\,\xr\,$:
\[ \Bxl:=\bmat e^{\Ixl(l)} & e^{\Ixl(r)} & e^{-\beta\frac{\muh+J}{4}\,+\,\Ixl(m)} \emat \;,\;
\Bxr:=\bmat e^{\Ixr(l)} \\ e^{\Ixr(r)} \\ e^{-\beta\frac{\muh+J}{4}\,+\,\Ixr(m)} \emat \;;\]
then to shorten the notation we set
\[ \bxl := \frac{1}{\sqrt{\la_1}}\,\Bxl\,\RE^{(1)} \;,\quad
\bxr := \frac{1}{\sqrt{\la_1}}\,\LE^{(1)}\Bxr \;. \]
%
Now define
\eq \label{eq: Poly defRL}
R_L \,:=\, \frac{Z_L}{\la_1^{|L|}\,\bxl^{\etaxl}\,\bxr^{\etaxr}}-\bxl^{\epsxl}\,\bxr^{\epsxr}
\eeq
and, using $\L$ as an abbreviation for $\L_\La(\cup_iS_i)$, rewrite the quantity $\prod_{L\in\L}Z_L$ by means of elementary algebraic tricks:
\[\begin{split}
\prod_{L\in\L} \frac{Z_L}{\la_1^{|L|}} \,&=\, \prod_{L\in\L} \left( \left( R_L +\, \bxl^{\epsxl}\,\bxr^{\epsxr} \right)\, \bxl^{\etaxl}\,\bxr^{\etaxr} \right) \\
&=\, \left(\prod_{L\in\L}\bxl^{\etaxl}\,\bxr^{\etaxr}\right)\ \sum_{\K\subseteq\L} \left(\prod_{L\in\K}\!R_L\right) \left(\prod_{L\in\L\setminus\K}\!\!\bxl^{\epsxl}\,\bxr^{\epsxr}\right) \;.
\end{split}\]
By identities \eqref{eq: Poly endptl}, \eqref{eq: Poly endptr} it holds
\begin{align*}
\prod_{L\in\L} \bxl^{\etaxl}\,\bxr^{\etaxr} \ &=\, \left(\prod_{x\in\partial_\l\La\setminus\cup_i\partial_\l S_i}\!\!\!\!\!\!b_{\l,x}\right)\, \left(\prod_{x\in\partial_\r\La\setminus\cup_i\partial_\r S_i}\!\!\!\!\!\!b_{\r,x}\right)  \\[6pt]
\prod_{L\in\L\setminus\K}\!\! \bxl^{\epsxl}\,\bxr^{\epsxr} \ &=\, \Bigg(\prod_{\substack{x\in(\cup_i\partial_\r^\ext S_i)\cap\La\\ x\notin\,\supp\K}}\!\!\!\!\!\!b_{\l,x}\Bigg)\; \Bigg(\prod_{\substack{x\in(\cup_i\partial_\l^\ext S_i)\cap\La\\ x\notin\,\supp\K}}\!\!\!\!\!\!b_{\r,x}\Bigg) \;;
\end{align*}
By substituting into the previous formula and thinking $\K=\{L_1,\dots,L_p\}$, we find out%
\footnote{In the first product on the r.h.s. of \eqref{eq: Poly ZLexpand} the shorten notation $b_{\l/\r,x}$ means: take $b_{\l,x}$ if $x\in\partial_\l\La$, take $b_{\r,x}$ if $x\in\partial_\r\La$; notice that $\partial_\l\La$ and $\partial_\r\La$ are disjoint for $N>1$. In the last product instead the shorten notation $b_{\r/\l,x}$ means: take $b_{\r,x}$ if $x\in\partial_\l^\ext S_i$ only, take $b_{\l,x}$ if $x\in\partial_\r^\ext S_i$ only, and take the product $b_{\r,x}\,b_{\l,x}$ in the case that $x$ belongs to both $\partial_\l^\ext S_i$ and $\partial_\r^\ext S_j$.}
\eq \label{eq: Poly ZLexpand} \begin{split}
\prod_{L\in\L} \frac{Z_L}{\la_1^{|L|}} \;=\; &
\Bigg(\prod_{x\in\partial_\v\La\setminus\cup_i\partial_\v S_i}\!\!\!\!\!\!\!\!b_{\l/\r,\,x}\Bigg) \,\cdot\\
& \cdot\, \sum_{p\geq0}\,\frac{1}{p!}\sum_{\substack{L_1,\dots,L_p\in\L\\ L_h\neq L_k\,\forall h\neq k}}
\left(\prod_{k=1}^p R_{L_k}\right)\, \Bigg(\prod_{\substack{x\in(\cup_i\partial_\v^\ext S_i)\cap\La\\ x\notin\cup_kL_k}}\!\!\!\!\!\!\!\!b_{\r/\l,\,x}\Bigg) \ .
\end{split} \eeq
Now substitute \eqref{eq: Poly ZLexpand} into \eqref{eq: Poly Z Si}, using also the fact that $|\La|=\sum_{i=1}^n|S_i|+\sum_{L\in\L_\La(\cup_iS_i)}|L|$, and obtain:
\eq  \label{eq: Poly Z Si Lk} \begin{split}
Z_\La^\h \;=\; & \la_1^{|\La|}\,\Bigg(\prod_{x\in\partial_\v\La}\!\!b_{\l/\r,\,x}\Bigg) \,\cdot\\
& \cdot \sum_{n\geq0}\, \frac{1}{n!}\!\!\!\! \sum_{\substack{S_1,\dots,S_n\,\in\,\S_\La\\ \,\dist(S_i,S_j)>1\,\forall i\neq j}} \prod_{i=1}^n  \left( \frac{e^{-\beta\left(\frac{\muh-\muv}{2}|S_i|\,+\,\frac{J}{2}|\partial_\h S_i|\right)}}{\la_1^{|S_i|}} \prod_{x\in\partial_\v\La\cap\partial_\v S_i}\frac{e^{-\beta\frac{J}{2}}}{b_{\l/\r,\,x}} \right) \,\cdot\\
& \cdot \sum_{p\geq0}\,\frac{1}{p!}\sum_{\substack{\,L_1,\dots,L_p\,\in\,\L_\La(\cup_i S_i)\\ L_k\neq L_h\,\forall k\neq h}}
\left(\prod_{k=1}^p R_{L_k}\right)\, \Bigg(\prod_{\substack{x\in(\cup_i\partial_\v^\ext S_i)\cap\La\\ x\notin\cup_kL_k}}\!\!\!\!\!\!\!\!b_{\r/\l,\,x}\Bigg) \ .
\end{split}\eeq
The next step is to partition $\bigcup_{i=1}^nS_i \,\cup\, \bigcup_{k=1}^pL_k$ into connected components as a sub-graph of $\widetilde\Z^2$, where $\widetilde\Z^2$ is the lattice obtained from $\Z^2$ by removing all the vertical bonds incident to the lines $L_k\,$:
\begin{gather*}
\bigcup_{i=1}^nS_i\,\cup\,\bigcup_{k=1}^pL_k \,=\, \bigcup_{t=1}^q\supp P_t \quad,\\[4pt]
P_t\in\P_\La\ \forall t \quad,\quad \dist_{\widetilde\Z^2}(\supp P_t,\supp P_s)>\!1\ \forall t\neq s 
\end{gather*}
where the family $\P_\La$ (yes, it is finally our family of polymers! see fig.\ref{fig3}) is defined by:
\eq \label{eq: Poly defP}
\P_\La :=\, \left\{ P\equiv\big((S_i)_{i\in I},(L_k)_{k\in K}\big) \;\middle|\; (S_i)_{i}\in\PS_\La\,,\ (L_k)_{k}\in\PL_{\La}(\cup_{i}S_i)  \right\} \;,
\eeq 
\eq \label{eq: Poly defPS}
(S_i)_{i\in I}\in\PS_\La \ \overset{\text{def}}{\Leftrightarrow}\ \begin{cases} \,0\leq|I|<\infty\\
\,S_i\in\S_\La\;\forall i\\ \,\dist_{\Z^2}(S_i,S_j)>1\;\forall i\neq j \;, \end{cases}
\eeq
\eq \label{eq: Poly defPL}
(L_k)_{k\in K}\in\PL_\La(\cup_{i\in I}S_i) \ \overset{\text{def}}{\Leftrightarrow}\ \begin{cases} \,0\leq|K|<\infty,\,|I|+|K|\geq1\\ \,L_k\in\L_\La(\cup_{i}S_i)\;\forall k\\ \,L_k\neq L_h\,\forall k\neq h\\ \,(\cup_{i}S_i)\cup(\cup_{k}L_k)\text{ connected in }\widetilde\Z^2 \;. \end{cases}
\eeq

\noindent The identity \eqref{eq: Poly Z Si Lk} now rewrites as
\eq \label{eq: Poly Z Pt}
Z_\La^\h \;=\;  C_\La\; \sum_{q\geq0}\,\frac{1}{q!} \sum_{P_1,\dots,P_q\in\P_\La}\, \prod_{t=1}^q\rho_\La(P_t)\ \prod_{t<s}\delta(P_t,P_s)
\eeq
by setting, for all $P,P'\in\P_\La$ with $P=\big((S_i)_{i\in I},(L_k)_{k\in K}\big)$,
\eq \label{eq: Poly defC}
C_\La \,:=\, \la_1^{|\La|}\, \prod\limits_{x\in\partial_\v\La}\!\!b_{\l/\r,\,x} \;,
\eeq
\eq \label{eq: Poly defrho} \begin{split}
\rho_\La(P) \,:=\; & \left( \frac{1}{|I|!}\, \prod_{i\in I}  \Bigg( \frac{e^{-\beta\left(\frac{\muh-\muv}{2}|S_i|\,+\,\frac{J}{2}|\partial_\h S_i|\right)}}{\la_1^{|S_i|}} \prod_{x\in\partial_\v\La\cap\partial_\v S_i}\frac{e^{-\beta\frac{J}{2}}}{b_{\l/\r,\,x}} \Bigg) \right) \cdot\\[2pt]
&\cdot \left( \frac{1}{|K|!}\, \prod_{k\in K} R_{L_k} \right)\,
\Bigg(\prod_{x\in(\bigcup_{i\in I}\!\partial_\v^\ext S_i)\cap\La \atop x\notin\,\bigcup_{k\in K}\!L_k}\!\!\!\!\!\!\!\!b_{\r/\l,\,x}\Bigg) \;,
\end{split} \eeq
\eq \label{eq: Poly defdelta}
\delta(P,P') \,:=\, \begin{cases} 
1 \,, & \text{if }\dist_{\widetilde\Z^2}(P,P')>1 \\
0 \,, & \text{otherwise} \end{cases} \;.
\eeq
The identity \eqref{eq: Poly Z Pt} finally shows that the partition function $Z_\La^\h\,$, up to a factor $C_\La$, admits a polymer representation of the form \eqref{eq: absCE Z}.

It is convenient to bound the polymer activity $\rho_\La$ by a simpler quantity. 
Using the proposition \ref{prop: 1D RL bounds} plus the lemmas \ref{lem: 1D endpt S}, \ref{lem: 1D endpt La} and the fact that $|\partial_\h S_i|\geq2$, one finds:
\eq \label{eq: Poly rhotilde}
\rho_\La(P) \,\leq\, \widetilde\rho(P) \,:=\, \left( \frac{1}{|I|!}\, \prod_{i\in I} e^{-\beta\left(\frac{\muh-\muv}{2}|S_i|\,+\,J\right)} \right) \left( \frac{1}{|K|!}\, \prod_{k\in K} e^{-m|L_k|}\,\ga_{L_k} \right)
\eeq
with the $\gamma_L$'s defined by the equation \eqref{eq: Poly gamma}.

\section{Convergence of the Cluster Expansion} \label{sec: CE}
In the previous section we rewrote our partition function $Z_\La^\h$ as a polymer partition function up to a factor $C_\La$ (see formula \eqref{eq: Poly Z Pt}).
In this section we will find a region of the parameters space $\muh,\muv,J$ where the condition \eqref{eq: absCE KPcondition} is verified by our model at low temperature, so that the general theorem \ref{thm: absCE main} about the convergence of the cluster expansion will apply to our case.

\begin{theorem} \label{thm: CE KPcond}
Assume that $J>0$, $\muh+J>0$ and $2\muv+5J<0\,$.
By choosing 
\eq \label{eq: CE a}
a(P):=\frac{m}{2}\,|\supp P| \quad\forall\,P\in\P_\La \eeq
the conditions
\begin{gather}
\label{eq: CE KPcond2}
\sum_{\substack{P\in\P_\La \\ \supp P\ni x}}\!\! \widetilde\rho(P)\,e^{a(P)} \,\leq\, \frac{m}{8} \quad\forall\,x\in\La \;,\\[4pt]
\label{eq: CE KPcond}
\sum_{\substack{ P\in\P_\La \\ \delta(P,P^*)=0}}\!\! \widetilde\rho(P)\,e^{a(P)} \,\leq\, a(P^*) \quad\forall\,P^*\!\in\P_\La
\end{gather}
hold true, provided that $\beta>\beta_0$ and $N>N_0(\beta)$ ($N$ is the minimum distance between two vertical components of $\partial\La$). Here $\beta_0>0$ depends on $\muh,\muv,J$ only, while $N_0(\beta)$ depends on $\beta,\muh,J$ only; they do not depend on $\La,\,P^*,\,x\,$.
\end{theorem}

\begin{corollary} \label{cor: CE}
Assume that $J>0$, $\muh+J>0$ and $2\muv+5J<0\,$. Suppose also that $\beta>\beta_0$ and $N>N_0(\beta)$.
Denote by $\CP_\La$ the set of clusters\footnote{As explained in the Appendix \ref{sec: absCE}, using the definition \eqref{eq: Poly defdelta} for $\delta$, a family of polymers $(P_1,\dots,P_q)$ is a \textit{cluster} iff $\cup_{t=1}^q\supp P_t$ is connected in $\widetilde\Z^2$.} composed by polymers of $\P_\La\,$.
Then the partition function \eqref{eq: Poly Z} rewrites as
\eq \label{eq: CE}
Z_\La^\h \,=\, C_\La\, \exp\bigg(\; \sumPLa\!\! U_\La\big((P_t)_t\big) \bigg)
\eeq
where we denote $\sum_{(P_t)_t\in\CP_\La}^* := \sum_{q\geq0}\frac{1}{q!}\sum_{(P_t)_{t=1}^q\in\CP_\La}$ and
\eq \label{eq: CE defU}
U_\La(P_1,\dots,P_q):= u(P_1,\dots,P_q)\, \prod_{t=1}^q\rho_\La(P_t) \;.
\eeq
Remind that $C_\La$ is defined by \eqref{eq: Poly defC}, $\rho_\La$ is defined by \eqref{eq: Poly defrho} and $u$ is defined by \eqref{eq: absCE u}, \eqref{eq: Poly defdelta}.
Furthermore for all $\mathscr E\subseteq\P_\La$ it holds
\eq \label{eq: CE clu-poly}
\sumPLaE \big|U_\La\big((P_t)_t\big)\big| \,\leq\, 
\sum_{\substack{P\in\P_\La \\ P\in\mathscr E}} |\rho_\La(P)|\,e^{a(P)}
\eeq
where $a$ is defined by \eqref{eq: CE a}.
\end{corollary}

\proof
The corollary follows from the general theory of cluster expansion (theorem \ref{thm: absCE main}), since $Z_\La^\h$ admits a polymer representation \eqref{eq: Poly Z Pt} and satisfies the Kotecky-Preiss condition (\eqref{eq: CE KPcond}, $|\rho_\La|\leq\widetilde\rho\,$). \qed
\endproof

For ease of reading, in the following of this section we will denote
\[  \sumS \,:=\, \sum_{n}\, \frac{1}{n!} \sum_{\,(S_i)_{i=1}^n\in\PS_\La} \ \quad\text{and}\qquad 
\sumL \,:=\, \sum_{p}\, \frac{1}{p!} \sum_{\,(L_k)_{k=1}^p\in\PL_\La(\cup_iS_i)} \]
where $\PS_\La$, $\PL_\La(\cup_iS_i)$ are the projections of the polymer set $P_\La$ defined in \eqref{eq: Poly defPS}, \eqref{eq: Poly defPL}.
The next lemmas provide the entropy estimates that will be needed in the proof of theorem \ref{thm: CE KPcond}.

\begin{lemma} \label{lem: CE entropy lines}
If $\cup_iS_i\neq\emptyset$, namely $n\geq1$, then
\eq \label{eq: CE entropy lines}
\sumL 1 \,\leq\, 4^{\sum_i|S_i|} \;.
\eeq
\end{lemma}
\proof
Fix $p\geq0$ and denote by $\PL^{(p)}_\La(\cup_iS_i)$ the set of $(L_k)_{k=1}^p\in\PL_\La(\cup_iS_i)$.\\
Given $(L_k)_{k=1}^p\in\PL^{(p)}_\La(\cup_iS_i)$, each line $L_k$ has at least one endpoint on $\cup_i\partial_\v^\ext S_i\,$, since $(\cup_iS_i)\cup(\cup_kL_k)$ have to be connected in $\widetilde\Z^2$.
Therefore the number of ways to choose each $L_k$ is at most $\sum_i|\partial_\v^\ext S_i| \leq 2\sum_i|S_i|\,$.
Since the $L_k$, $k=1,\dots,p$, must be all distinct, it follows that
\[ \left|\PL^{(p)}_\La(\cup_iS_i)\right| \,\leq\, \big(2\sum\nolimits_i|S_i|\big)\,\big(2\sum\nolimits_i|S_i|-1\big)\,\cdots\,\big(2\sum\nolimits_i|S_i|-p+1\big) \;.\]
Therefore
\[ \sumL 1 \,=\, \sum_{p}\, \frac{1}{p!}\, \left|\PL^{(p)}_\La(\cup_iS_i)\right| \,\leq\, \sum_{p} {2\sum_i|S_i| \choose p} \,=\, 2^{2\sum_i|S_i|} \;. \] \qed
\endproof

\begin{lemma} \label{lem: CE entropy S,x}
Let $x\in\Z^2$. For all $s\geq2$
\eq \label{eq: CE entropy S,x}
\card\left\{S\subset\Z^2\text{ connected} \;\middle|\; |S|=s\,,\ S\ni x \right\} \,\leq\, \frac{16}{3}\,4^{4s} \;.
\eeq
\end{lemma}
\proof
Given a connected graph $G$ and one of its vertices $x$, there exists a walk in $G$ that starts from $x$ and crosses each edge exactly twice%
\footnote{This can be easily proven by induction on the number of edges.}.
Therefore
\[\begin{split}
& \card\left\{S\subset\Z^2\text{ connected} \;\middle|\; |S|=s\,,\ S\ni x \right\} \,\leq\\
&\leq \sum_{e=s-1}^{2s} \card\left\{S\text{ connected sub-graph of }\Z^2 \;\middle|\; |\text{edges of }S|=e\,,\ S\ni x \right\} \\
&\leq \sum_{e=s-1}^{2s} \card\left\{\text{walks in $\Z^2$ that start from $x$ and have lenght $2e$} \right\} \\
&\leq \sum_{e=s-1}^{2s} 4^{2e} \;\leq\; \frac{4^{4s+2}}{3} \;.
\end{split}\] \qed
\endproof

\begin{lemma} \label{lem: CE entropy S,A}
Let $A\subset\Z^2$ finite. For all $s\geq2$, $1\leq d<\infty$ 
\eq \label{eq: CE entropy S,A}
\card\left\{S\subset\Z^2\text{ connected} \;\middle|\; |S|=s\,,\ \dist_\h(S,A)=d \right\} \,\leq\, \frac{32}{3}\,|A|\,4^{4s} \;.
\eeq
Here $\dist_\h(S,A):=\inf_{x\in S,\,y\in A}\dist_\h(x,y)$ and the horizontal distance between $x=(x_\h,x_\v),\,y=(y_\h,y_\v)\in\Z^2$ is defined as
\eq \label{eq: CE defdisth}
\dist_\h(x,y) \,:=\, \begin{cases}\ |x_\h-y_\h| & \text{if }x_\v=y_\v \\\ +\infty & \text{if }x_\v\neq y_\v \end{cases} \;.
\eeq
\end{lemma}
\proof
Observe that $\dist_\h(S,A)=d$ if and only if there exists a horizontal line $L$, $|L|=d+1$, having one endpoint on $\partial_\v A$ and the other one on $\partial_\v S\,$. Therefore:
\[\begin{split} 
& \card\left\{S\subset\Z^2\text{ connected} \;\middle|\; |S|=s\,,\ \dist_\h(S,A)=d \right\} \,\leq\\
&\leq \sum_{\substack{L\text{ horiz.}\,\text{line, }|L|=d+1, \\ \partial_\v A\,\ni\text{ one endpt.}\,\text{of $L$}}}\!\!\!\! \card\left\{S\subset\Z^2\,\text{\small connected} \;\middle|\; |S|=s\,,\ \partial_\v S\ni\text{\small other endpt.}\,\text{\small of $L$} \right\} \\
&\leq\, 2|\partial_\v A|\,\ \card\left\{S\subset\Z^2\text{ connected} \;\middle|\; |S|=s\,,\ S\ni 0 \right\} 
\,\leq\, 2|A|\ \frac{16}{3}\,4^{4s} \;.
\end{split}\]
For the last inequality we have used the lemma \ref{lem: CE entropy S,x}. \qed
\endproof

\begin{lemma} \label{lem: CE I entropy Sfam}
Let $n\geq1\,$. Let $\T$ be a tree over the vertices $\{1,\dots,n\}\,$. Let $s_i\geq2$ for all $i=1,\dots,n\,$ and $d_{ij}\geq2$ for all $(i,j)\in\T$.\\
Then given $A\subset\Z^2$ and $1\leq d<\infty$
\eq \label{eq: CE entropy treeA} \begin{split}
& \card\,\big\{ (S_i)_{i=1}^n\in\PS_\La \ \big|\; \dist_\h(S_1,A)=d\,,\ |S_i|=s_i\,\forall i\,,\\ &\qquad\qquad\qquad\qquad\qquad \dist_\h(S_i,S_j)=d_{ij}\,\forall(i,j)\!\in\!\T  \big\} \,\leq\\[2pt]
&\leq\, |A|\, \prod_{i=1}^n\left(\frac{32}{3}\,4^{4s_i}\,s_i^{\deg_{\T}(i)}\right) \;;
\end{split} \eeq
while given $x\in\Z^2$
\eq \label{eq: CE entropy treex} \begin{split}
& \card\,\big\{ (S_i)_{i=1}^n\in\PS_\La \ \big|\; S_1\ni x\,,\ |S_i|=s_i\,\forall i\,,\\ &\qquad\qquad\qquad\qquad\qquad \dist_\h(S_i,S_j)=d_{ij}\,\forall(i,j)\!\in\!\T  \big\} \,\leq\\
&\leq\, \prod_{i=1}^n\left(\frac{32}{3}\,4^{4s_i}\,s_i^{\deg_{\T}(i)}\right) \;.
\end{split} \eeq
Here $\deg_{\T}(i)$ denotes the degree of the vertex $i$ in the tree $\T$.
\end{lemma}

\proof
Let start by proving the inequality \eqref{eq: CE entropy treeA} by induction on $n$.
If $n=1$, then the tree $\T$ is trivial and \eqref{eq: CE entropy treeA} is already provided by the lemma \ref{lem: CE entropy S,A}.
Now let $n\geq2$, assume that \eqref{eq: CE entropy treeA} holds for at most $n-1$ vertices and prove it for $n\,$. 
It is convenient to think that the tree $\T$ is rooted at the vertex $1$ and denote by $j\leftarrow i$ the relation \virg{vertex $j$ is son of vertex $i$ in $\T$} and by $\T(i)$ the sub-tree of $\T$ induced by the vertex $i$ together with its descendants. Then, denoting by $N_{\T,1}\big(A,d;(s_i)_{i\in\T},(d_{ij})_{(i,j)\in\T}\big)$ the cardinality on the l.h.s. of \eqref{eq: CE entropy treeA}, it holds
\[\begin{split}
& N_{\T,1}\!\left(A,d;(s_i)_{i\in\T},(d_{ij})_{(i,j)\in\T}\right) \,=\\
& = \sum_{\substack{S_1\in\S_\La,\;|S_1|=s_1 \\ \dist_\h(S_1,A)=d}}\, \prod_{v\leftarrow1}\, N_{\T(v),v}\!\left(S_1,d_{1v};(s_i)_{i\in\T(v)},(d_{ij})_{(i,j)\in\T(v)}\right) \;.\end{split}\]
Since $\T(v)$ has at most $n-1$ vertices, the induction hypothesis gives
\[ N_{\T(v),v}\!\left(S_1,d_{1v};(s_i)_{i\in\T(v)},(d_{ij})_{(i,j)\in\T(v)}\right) \,\leq\, s_1 \prod_{i\in\T(v)}\left(\frac{32}{3}\,4^{4s_i}\,s_i^{\deg_{\T(v)}(i)}\right) \;.\]
Then by substituting in the previous identity, bounding $\deg_{\T(v)}(i)$ by $\deg_\T(i)$ and using the lemma \ref{lem: CE entropy S,A}, one obtains:
\[ N_{\T,1}\!\left(A,d;(s_i)_{i\in\T},(d_{ij})_{(i,j)\in\T}\right) \,\leq\,
|A|\, \prod_{i\in\T} \left(\frac{32}{3}\,4^{4s_i}\,s_i^{\deg_{\T}(i)}\right) \;.\]
This concludes the proof of \eqref{eq: CE entropy treeA}.

\noindent In order to prove the inequality \eqref{eq: CE entropy treex}, denote by $N'_{\T,1}\big(x;(s_i)_{i\in\T},(d_{ij})_{(i,j)\in\T}\big)$ the cardinality on the l.h.s. of \eqref{eq: CE entropy treex} and observe that
\[\begin{split}
& N'_{\T,1}\!\left(x;(s_i)_{i\in\T},(d_{ij})_{(i,j)\in\T}\right) \,=\\
& \sum_{\substack{S_1\in\S_\La,\;|S_1|=s_1\\ S_1\ni x}}\, \prod_{v\leftarrow1}\, N_{\T(v),v}\!\left(S_1,d_{1v};(s_i)_{i\in\T(v)},(d_{ij})_{(i,j)\in\T(v)}\right) \;.\end{split}\]
By \eqref{eq: CE entropy treeA} we already know that
\[ N_{\T(v),v}\!\left(S_1,d_{1v};(s_i)_{i\in\T(v)},(d_{ij})_{(i,j)\in\T(v)}\right) \,\leq\, s_1 \prod_{i\in\T(v)}\left(\frac{32}{3}\,4^{4s_i}\,s_i^{\deg_{\T(v)}(i)}\right) \;.\]
Then by substituting in the previous identity, bounding $\deg_{\T(v)}(i)$ by $\deg_\T(i)$ and using the lemma \ref{lem: CE entropy S,x}, one obtains:
\[ N'_{\T,1}\!\left(x;(s_i)_{i\in\T},(d_{ij})_{(i,j)\in\T}\right) \,\leq\,
\prod_{i\in\T} \left(\frac{32}{3}\,4^{4s_i}\,s_i^{\deg_{\T}(i)}\right) \;,\]
which proves \eqref{eq: CE entropy treex}. \qed
\endproof

\proof[of the theorem \ref{thm: CE KPcond}]
According to the definition \eqref{eq: Poly defdelta}, the condition $\delta(P,P^*)=0$ implies that $\supp P \cap [\supp P^*]_1 \neq \emptyset\,$, where $[A]_1:=\{x\in\Z^2\,|\,\dist_{\Z^2}(x,A)\leq1\}\,$.
Therefore
\[\begin{split}
\sum_{P\in\P_\La\atop\delta(P,P^*)=0}\!\! \widetilde\rho(P)\,e^{a(P)} \,&\leq\, 
\sum_{x\in[\supp P^*]_1} \sum_{P\in\P_\La\atop\supp P\ni x}\!\! \widetilde\rho(P)\,e^{a(P)} \\
&\leq\, 4\,|\supp P^*|\; \max_{x\in\La} \sum_{P\in\P_\La\atop\supp P\ni x}\!\! \widetilde\rho(P)\,e^{a(P)} \;.
\end{split}\]
Thus, by choosing $a(P):=\frac{m}{2}|\supp P|$ for all $P\in\P_\La$, the inequality \eqref{eq: CE KPcond} will be a consequence of \eqref{eq: CE KPcond2}. 

We have to prove the inequality \eqref{eq: CE KPcond2}. 
It is worth to write down explicitly the quantity we will work with (see the definitions \eqref{eq: Poly rhotilde} and \eqref{eq: CE a}):
\[ \widetilde\rho(P)\,e^{a(P)} \,=\,
\Bigg( \frac{1}{n!}\, \prod_{i=1}^n e^{-\left(\beta\frac{\muh-\muv}{2}-\frac{m}{2}\right)|S_i|-\beta J} \Bigg) \Bigg( \frac{1}{p!}\, \prod_{k=1}^p e^{-\frac{m}{2}|L_k|}\,\ga_{L_k} \Bigg) \]
for all $P\in\P_\La$, $P=\big((S_i)_{i=1}^n,(L_k)_{k=1}^p\big)\,$.
Notice that if $\supp P\ni x$, the site $x$ may belong either to a region $S_i$ or to a line $L_k$; hence we can split the sum on the l.h.s. of \eqref{eq: CE KPcond2} into two parts:
\eq
\sum_{P\in\P_\La\atop\supp P\ni x}\!\! \widetilde\rho(P)\,e^{a(P)} \,=\, \Sigma_1 \,+\, \Sigma_2
\eeq
with
\begin{gather}
\Sigma_1 \,:=\, \sumSx \left(\prod_{i} e^{-\left(\beta\frac{\muh-\muv}{2}-\frac{m}{2}\right)|S_i|-\beta J}\right)\, \sumL \prod_{k} e^{-\frac{m}{2}|L_k|}\,\ga_{L_k} \;\;\\[4pt]
\Sigma_2 \,:=\, \sumS \left(\prod_{i}e^{-\left(\beta\frac{\muh-\muv}{2}-\frac{m}{2}\right)|S_i|-\beta J}\right)\, \sumLx \prod_{k} e^{-\frac{m}{2}|L_k|}\,\ga_{L_k}  \;.
\end{gather}
During all the proof $o(1)$ will denote any function $\omega=\omega(\beta,\muh,J)$ such that $\omega\to0$ as $\beta\to\infty$ and $\omega$ depends only on $\beta,\,\muh,\,J$ (in particular it does not depend on the choices of $\La\subset\Z^2$, $x\in\Z^2$, $P\in\P_\La$).\\

\textit{\textbf{I}. Study of the term $\Sigma_1$.}

We fix a family of regions $(S_i)_{i=1}^n$ that contains the point $x$; we also assume that $\PL_\La(\cup_iS_i)$ is non-empty, otherwise the contribution to $\Sigma_1$ is zero.
By the lemma \ref{lem: CE entropy lines} it holds
\eq \label{eq: CE I bound Lsum}
\sumL\, \prod_{k} e^{-\frac{m}{2}|L_k|}\,\ga_{L_k} \;\leq\;
4^{\sum_i|S_i|}\, \max_{(L_k)_k}\,\prod_{k}e^{-\frac{m}{2}|L_k|}\,\ga_{L_k} \eeq
where the maximum is taken over all $(L_k)_{k}\in\PL_\La(\cup_iS_i)\,$.
The factor $\ga_{L_k}$ can take two values (see formula \eqref{eq: Poly gamma}), both smaller than $1$ for $\beta$ sufficiently large (uniformly with respect to $L_k$), since each line $L_k$ must have at least one endpoint on $\cup_i\partial_\v^\ext S_i\,$ to ensure that $(\cup_iS_i)\cup(\cup_kL_k)$ is connected in $\widetilde\Z^2$.

Obviously $n\geq1$ in order for $\cup_{i=1}^n S_i$ to contain the point $x$. It is convenient to consider separately the case $n=1$ and the case $n\geq2\,$:
\[ \Sigma_1 \,=\, \Sigma_1' + \Sigma_1'' \;.\]
The case $n=1$ is easy to deal with, simply by bounding the r.h.s. of \eqref{eq: CE I bound Lsum} by $4^{|S|}$ and using the lemma \ref{lem: CE entropy S,x}. Precisely:
\eq \label{eq: CE I n1} \begin{split}
\Sigma_1' \,&:= \sum_{\substack{S\in\S_\La\\ S\ni x}} e^{-\left(\beta\frac{\muh-\muv}{2}-\frac{m}{2}\right)|S|-\beta J}\ \sumL\prod_k e^{-\frac{m}{2}|L_k|}\gamma_{L_k} \\
&\leq\, \sum_{\substack{S\in\S_\La\\ S\ni x}} e^{-\left(\beta\frac{\muh-\muv}{2}-\frac{m}{2}\right)|S|-\beta J}\; 4^{|S|} \\
&\leq\, \sum_{s\geq2\atop\text{even}} \frac{16}{3}\,4^{4s}\, e^{-\left(\beta\frac{\muh-\muv}{2}-\frac{m}{2}\right)s-\beta J}\, 4^s \\
&=\, \frac{16}{3}\,4^{10}\,e^{-\beta\,(\muh-\muv+J)}\,(1+o(1)) \;.
\end{split} \eeq

Now assume $n\geq2$. Fix a family of lines $(L_k)_{k=1}^p\in\PL_\La(\cup_iS_i)\,$. We can consider the graph $G\equiv G\big((S_i)_i,(L_k)_k\big)$ with vertices $i\in\{1,\dots,n\}$ and edges $k\in\{1,\dots,p\}\,$: the edge $k$ joins the two vertices $i,j$ iff the line $L_k$ has one endpoint on $\partial_\v^\ext S_i$ and the other one on $\partial_\v^\ext S_j$. In the graph $G$ there can be multiple edges, loops and pseudo-edges with a single endpoint. The graph $G$ is connected (it follows from definition \ref{eq: Poly defPL}), hence $G$ admits at least one spanning sub-tree $\T$.
And clearly, since each factor $e^{-\frac{m}{2}|L_k|}\,\ga_{L_k}$ is smaller than $1$,
\[ \prod_{k=1}^p e^{-\frac{m}{2}|L_k|}\,\ga_{L_k} \,\leq\,
\prod_{k\in\T} e^{-\frac{m}{2}|L_k|}\,\ga_{L_k}  \,\leq
\prod_{(i,j)\in\T}e^{-\frac{m}{2}\left(\dist_\h(S_i,S_j)-1\right)}\,\ga_{S_i,S_j} \]
where $\gamma_{S,S'}:=\big( \frac{1}{2}e^{-\beta J}+e^{-\beta\frac{\muh+J}{2}(\dist_\h(S,S')-1)} \big)\,(1+o(1))\,$. 
Therefore:
\eq \label{eq: CE I bound Lmax}
\max_{(L_k)_k}\,\prod_{k}e^{-\frac{m}{2}|L_k|}\,\ga_{L_k} \,\leq\,
\max_{\substack{\T\text{ tree over}\\ \{1,\dots,n\}}} \prod_{(i,j)\in\T}e^{-\frac{m}{2}\left(\dist_\h(S_i,S_j)-1\right)}\,\ga_{S_i,S_j}
\eeq
%
Now using \eqref{eq: CE I bound Lsum} and \eqref{eq: CE I bound Lmax} we can bound $\Sigma_1''\,$:
\eq \label{eq: CE I bound Sigma1} \begin{split}
\Sigma_1'' \,&:=\,
\sum_{n\geq2} \frac{1}{n!} \sum_{\substack{(S_i)_{i=1}^n\\ \cup_i S_i\ni x}} \left(\prod_{i=1}^n e^{-\left(\beta\frac{\muh-\muv}{2}-\frac{m}{2}\right)|S_i|-\beta J}\right)\, \sumL \prod_{k} e^{-\frac{m}{2}|L_k|}\,\ga_{L_k} \\[2pt]
&\leq\, \sum_{n\geq2} \sum_{\,\substack{\T\!\text{ tree over}\\ \{1,\dots,n\}}} \frac{1}{n!} \sum_{\substack{(S_i)_{i=1}^n\\ \cup_iS_i\ni x}}\, \left(\prod_{i=1}^n e^{-\left(\beta\frac{\muh-\muv}{2}-\frac{m}{2}-\log4\right)|S_i|-\beta J}\right) \,\cdot\\ & \qquad\qquad\qquad\qquad\qquad\qquad\qquad \cdot\prod_{(i,j)\in\T}\!\!e^{-\frac{m}{2}\left(\dist_\h(S_i,S_j)-1\right)}\,\ga_{S_i,S_j}
\end{split} \eeq
where in the sums we keep implicit that $(S_i)_{i=1}^n\in\PS_\La$. 

\noindent Substitute into \eqref{eq: CE I bound Sigma1} the entropy bound\footnote{The families of regions $(S_i)_{i=1}^n$ such that $\dist_\h(S_i,S_j)=\infty$ for at least one edge $(i,j)\in\T$ give zero contribution to the sum, therefore we do not need to worry about them.} \eqref{eq: CE entropy treex}. Since $\cup_iS_i\ni x$, but not necessarily $S_1\ni x$, an extra factor $n$ appears. Moreover observe that $|S_i|$ is even and $\geq2$ (see the definition \eqref{eq: Poly defS}) and $\dist_\h(S_i,S_j)\geq2\,$. Then:
\eq \label{eq: CE I boundentropy Sigma1} \begin{split}
\Sigma_1'' \,\leq\; &
\sum_{n\geq2} \sum_{\,\substack{\T\!\text{ tree over}\\ \{1,\dots,n\}}} \frac{n}{n!} \sum_{\substack{(s_i)_{i=1,\dots,n}\\ s_i\!\text{ even }\geq2}} \sum_{\substack{(d_{ij})_{ij\in\T}\\ d_{ij}\geq2}} \left(\prod_{i=1}^n \frac{32}{3}\,4^{4s_i}\,s_i^{\deg_\T(i)}\right) \cdot\\
& \cdot\,\left(\prod_{i=1}^n e^{-\left(\beta\frac{\muh-\muv}{2}-\frac{m}{2}-\log4\right)s_i-\beta J}\right) \prod_{(i,j)\in\T}\!\!e^{-\frac{m}{2}\left(d_{ij}-1\right)}\,\ga_{d_{ij}}
\end{split} \eeq
where $\ga_{d}:=\big( \frac{1}{2}e^{-\beta J}+e^{-\beta\frac{\muh+J}{2}\,(d-1)} \big)\,(1+o(1))\,$. 

\noindent Given $n\geq2$ and $\delta_1,\dots,\delta_n\geq1$, the number of trees $\T$ over the vertices $\{1,\dots,n\}\,$ with given degrees $\deg_\T(i)=\delta_i\,$ $\forall i=1,\dots,n$ is exactly\footnote{This is an improvement of the well-known Cayley's formula.}
\[ \frac{(n-2)!}{(\delta_1-1)!\cdots(\delta_n-1)!} \]
if $\sum_{i=1}^n(\delta_i-1)=n-2$ and zero otherwise.
Furthermore the number of edges of $\T$ is $n-1$.
Therefore the bound \eqref{eq: CE I boundentropy Sigma1} leads to
\eq \label{eq: CE I boundrearr Sigma1} \begin{split}
\Sigma_1'' \,\leq\,
\sum_{n\geq2}\; & \Bigg( \frac{32}{3}\,e^{-\beta J}\, \sum_{s\geq2\atop \text{even}} e^{-\left(\beta\frac{\muh-\muv}{2}-\frac{m}{2}-5\log4\right)s}\, \sum_{\delta\geq1}\frac{s^{\delta}}{(\delta-1)!} \Bigg)^{\!n} \,\cdot\\
& \cdot \Bigg(\sum_{d\geq2}e^{-\frac{m}{2}\left(d-1\right)}\,\ga_{d}\Bigg)^{\!n-1} \;.
\end{split} \eeq
The sum over $s$ gives, as $\beta\to\infty$,
\eq \label{eq: CE I sums} \begin{split}
&\sum_{s\geq2\atop\text{even}} e^{-\left(\beta\frac{\muh-\muv}{2}-\frac{m}{2}-5\log4\right)s}\, \sum_{\delta\geq1}\frac{s^{\delta}}{(\delta-1)!} \,=\\
&=\, \sum_{s\geq2\atop\text{even}} s\,e^{-\left(\beta\frac{\muh-\muv}{2}-\frac{m}{2}-5\log4-1\right)s}  \,=\, 2\,e^2\,4^{10}\,e^{-\beta(\muh-\muv)}\,(1+o(1)) \;.
\end{split} \eeq
The sum over $d$ gives, as $\beta\to\infty$,
\eq \label{eq: CE I sumd} \begin{split}
& \sum_{d\geq2}e^{-\frac{m}{2}\left(d-1\right)}\,\ga_{d} \,=\\
&= \Bigg( \sum_{d\geq2}e^{-\frac{m}{2}\left(d-1\right)}\, \frac{e^{-\beta J}}{2} \;+\; \sum_{d\geq2}e^{-\frac{m}{2}(d-1)}\,e^{-\beta\frac{\muh+J}{2}(d-1)} \Bigg)\, (1+o(1)) \\
&= \Big(  \frac{1}{1-e^{-\frac{m}{2}}}\, \frac{e^{-\beta J}}{2} \;+\; o(1) \Big)\, (1+o(1)) \;=\;
e^{\beta\frac{\muh+J}{2}}\, (1+o(1))
\end{split} \eeq
where we used the fact that $1-e^{-\frac{m}{2}} \,=\, \frac{1}{2}\,e^{-\beta\frac{\muh+3J}{2}}\,(1+o(1))$ (see lemma \ref{lem: 1D eigenval ratio}).
Substituting \eqref{eq: CE I sums}, \eqref{eq: CE I sumd} into \eqref{eq: CE I boundrearr Sigma1}, one obtains
\eq \label{eq: CE I boundclose Sigma1}
\Sigma_1'' \,\leq\, \sum_{n\geq2}\, \Bigg( \frac{2^{26}e^2}{3}\,e^{-\beta(\muh-\muv)+\beta\frac{\muh+J}{2}}\,(1+o(1)) \Bigg)^{\!n} \,  e^{-\beta\frac{\muh+J}{2}}\, (1+o(1)) \;.\eeq
Assume $\boldsymbol{\muh-\muv>\frac{\muh+J}{2}}\,$. Then for $\beta$ sufficiently large \eqref{eq: CE I boundclose Sigma1} becomes:
\eq \label{eq: CE I boundfinal Sigma1} \begin{split}
\Sigma_1'' \,&\leq\,
\left( \frac{2^{26}e^2}{3}\, e^{-\beta(\muh-\muv)+\beta\frac{\muh+J}{2}} \right)^2\; e^{-\beta\frac{\muh+J}{2}}\,(1+o(1)) \\[2pt]
&=\, \frac{2^{52}e^4}{9}\; e^{-\beta\,2(\muh-\muv)+\beta\,\frac{\muh+J}{2}}\, (1+o(1)) \;.
\end{split} \eeq

\textit{\textbf{II.} Study of the term $\Sigma_2$.}

The ideas are not far from those already seen for $\Sigma_1$.
We fix a family of regions $(S_i)_{i=1}^n$ and we assume that there exists $(L_k)_k\in\PL_\La(\cup_i S_i)$ such that $\cup_kL_k\ni x\,$, otherwise the contribution to $\Sigma_2$ is zero. Clearly the line $L^{x}\in\L_\La(\cup_iS_i)$ that contains $x$ is unique.
It is convenient to consider separately four cases:
\[ \Sigma_2 \,=\, \Sigma_2' + \Sigma_2'' + \Sigma_2''' +\Sigma_2'''' \;.\]
In $\Sigma_2'$ we assume $n=0$, namely $\cup_iS_i=\emptyset\,$; then $L^{x}$ have to be a maximal horizontal line of $\La\,$.
In $\Sigma_2''$ we assume $n=1$, namely there is a unique region $S$ and $L^{x}$ may have one endpoint on $\partial_\v^\ext S$ and one on $\partial_\v\La$ or both on $\partial_\v^\ext S\,$. 
In $\Sigma_2'''$ we assume $n\geq2$ and $L^{x}$ has one endpoint on $\cup_i\partial_\v^\ext S_i$ and one on $\partial_\v\La\,$ or both on the same $\partial_\v^\ext S_i$.
In $\Sigma_2''''$ we assume $n\geq2$ and $L^{x}$ has one endpoint on $\partial_\v^\ext S_i$ and one on $\partial_\v^\ext S_j$ with $i\neq j$.

The case $n=0$ is easy to deal with. Indeed, since the unique $(L_k)_k\in\PL_\La(\emptyset)$ such that $\cup_kL_k\ni x$ is the singleton $(L^{x})\,$,
\eq \label{eq: CE II n0}
\Sigma_2' \,:= \sumLx \prod_k e^{-\frac{m}{2}|L_k|}\, \gamma_{L_k}
\,=\, e^{-\frac{m}{2}|L^{x}|}\, \gamma_{L^{x}} 
\,\leq\, e^{-\frac{m}{2}N}\, (1+o(1))  \eeq
where $N$ denotes the minimum distance between two different vertical components of $\partial\La\,$. 

When $n\geq1$, by the lemma \ref{lem: CE entropy lines} it holds:
\eq \label{eq: CE II bound Lsum}
\sumLx\, \prod_{k} e^{-\frac{m}{2}|L_k|}\,\ga_{L_k} \;\leq\;
4^{\sum_i|S_i|}\, \max_{\substack{(L_k)_k\\ \cup_k L_k\ni x}}\,\prod_{k}e^{-\frac{m}{2}|L_k|}\,\ga_{L_k} \eeq
where it is implicit in the notation that $(L_k)_{k}\in\PL_\La(\cup_iS_i)\,$.
The factor $\ga_{L_k}$ can take two values (see formula \eqref{eq: Poly gamma}), both smaller than $1$ for $\beta$ sufficiently large (uniformly with respect to $L_k$), since each line $L_k$ must have at least one endpoint on $\cup_i\partial_\v^\ext S_i\,$. 

\noindent Now the case $n=1$ is also easy to deal with. Indeed, by bounding the r.h.s. of \eqref{eq: CE II bound Lsum} by $4^{|S|}\,e^{-\frac{m}{2}|L^x|}\,\ga_{L^x}$, one obtains:
\eq \begin{split}
\Sigma_2'' \,&:= \sum_{S\in\S_\La} e^{-\left(\beta\frac{\muh-\muv}{2}-\frac{m}{2}\right)|S|}\, \sumLx \prod_k e^{-\frac{m}{2}|L_k|}\, \gamma_{L_k} \\
&\leq\, \sum_{S\in\S_\La} e^{-\left(\beta\frac{\muh-\muv}{2}-\frac{m}{2}\right)|S|}\ 4^{|S|}\, e^{-\frac{m}{2}|L^{x}|}\,\ga_{L^{x}} \;;
\end{split} \eeq
then observe that $|L^{x}|\geq\dist_\h(S,x)$ and use the lemma \ref{lem: CE entropy S,A} for the entropy:
\eq \begin{split}
\Sigma_2'' \,&\leq\, \sum_{s\geq2\atop\text{even}}\, \sum_{d\geq1}\, \frac{32}{3}\,4^{4s}\, e^{-\left(\beta\frac{\muh-\muv}{2}-\frac{m}{2}\right)s}\, 4^s\, e^{-\frac{m}{2}d}\,\ga_d 
\end{split} \eeq
where $\gamma_d := \Big(\frac{e^{-\beta\frac{J}{2}}}{\sqrt2}+e^{-\beta\frac{\muh+J}{2}d}\Big)(1+o(1))\,$; 
finally compute the geometric series in $s,d$, and use $1-e^{-\frac{m}{2}} = \frac{1}{2}e^{-\beta\frac{\muh+3J}{2}}\,(1+o(1))$ (see lemma \ref{lem: 1D eigenval ratio}) to obtain:
\eq \label{eq: CE II n1} \begin{split}
\Sigma_2'' \,&\leq\, \frac{32}{3}\, \Big(4^{10}\, e^{-\beta(\muh-\muv)}\Big)\, \Big(\frac{1}{1-e^{-\frac{m}{2}}}\,\frac{e^{-\beta\frac{J}{2}}}{\sqrt 2}\,+\,o(1)\Big)\, (1+o(1)) \\[2pt] 
&=\, \frac{2^{25}\sqrt2}{3}\; e^{-\beta\,(\muh-\muv)+\beta\,\frac{\muh+2J}{2}}\,(1+o(1)) \;.
\end{split} \eeq

Assume now $n\geq2$ and that $L^x$ has one endpoint on $\cup_i\partial_\v^\ext S_i$ and the other one on $\partial^\ext\La$ or both endpoints on the same $\partial_\v^\ext S_i$. By introducing an extra factor $n$ we may assume that one endpoint is on $\partial_\v^\ext S_1$.
Fix a family of lines $(L_k)_{k=1}^p\in\PL_\La(\cup_iS_i)$ such that $\cup_kL_k\ni x$ (namely there is a $k$ such that $L_k=L^{x}$). We can introduce the graph $G\equiv G\big((S_i)_i,(L_k)_k\big)$ with vertices $i\in\{1,\dots,n\}$ and edges $k\in\{1,\dots,p\}\,$: the edge $k$ joins the two vertices $i,j$ iff the line $L_k$ has one endpoint on $\partial_\v^\ext S_i$ and the other one on $\partial_\v^\ext S_j$. The graph $G$ is connected, hence $G$ admits at least one spanning sub-tree $\T$. Notice that the line $L^{x}$ is not part of this tree.
Hence, since each factor $e^{-\frac{m}{2}|L_k|}\,\ga_{L_k}$ is smaller than $1$,
\[\begin{split} 
\prod_{k=1}^p e^{-\frac{m}{2}|L_k|}\,\ga_{L_k} \,&\leq\,
e^{-\frac{m}{2}|L^{x}|}\,\ga_{L^{x}}\, \prod_{k\in\T} e^{-\frac{m}{2}|L_k|}\,\ga_{L_k}  \\
&\leq\, e^{-\frac{m}{2}\dist_\h(S_1,x)}\,\ga_{S_1,x}\, \prod_{(i,j)\in\T}e^{-\frac{m}{2}\left(\dist_\h(S_i,S_j)-1\right)}\,\ga_{S_i,S_j} 
\end{split}\]
where $\ga_{S,x}:=\big(\frac{1}{\sqrt2}e^{-\beta\frac{J}{2}}+e^{-\beta\frac{\muh+J}{2}\dist_\h(S,x)}\big)\,(1+o(1))$ and $\ga_{S,S'}:=\big( \frac{1}{2}e^{-\beta J}+e^{-\beta\frac{\muh+J}{2}(\dist_\h(S,S')-1)} \big)\,(1+o(1))\,$. 
Therefore:
\eq \label{eq: CE II bound Lmax} \begin{split}
&\max_{\substack{(L_k)_k,\, \cup_k\!L_k\ni x,\\ L^{x}\!\text{ from }\partial_\v^\ext S_1\text{ to }\partial_\v\La\\ \text{ or from a }\partial_\v^\ext S_1\!\text{ to itself}}}\,\prod_{k}e^{-\frac{m}{2}|L_k|}\,\ga_{L_k} \,\leq\\[4pt]
&\leq\, e^{-\frac{m}{2}\dist_\h(S_1,x)}\,\ga_{S_1,x}\; \max_{\substack{\T\text{ tree over}\\ \{1,\dots,n\}}} \prod_{(i,j)\in\T}e^{-\frac{m}{2}\left(\dist_\h(S_i,S_j)-1\right)}\,\ga_{S_i,S_j} \;.
\end{split} \eeq
Now using \eqref{eq: CE II bound Lsum} and \eqref{eq: CE II bound Lmax} we can bound $\Sigma_2'''$:
\eq \label{eq: CE II bound Sigma23} \begin{split}
\Sigma_2''' \,&:=
\sum_{n\geq2} \frac{1}{n!} \sum_{(S_i)_{i=1}^n} \left(\prod_{i} e^{-\left(\beta\frac{\muh-\muv}{2}-\frac{m}{2}\right)|S_i|-\beta J}\right) \cdot\\ &\qquad\qquad\qquad\qquad \cdot \sideset{}{^*}\sum_{\substack{(L_k)_k,\,\cup_kL_k\ni x,\\ L^{x}\!\text{ from }\cup_i\partial_\v^\ext S_i\text{ to }\partial_\v\La\\ \text{ or from a }\partial_\v^\ext S_i\!\text{ to itself}}} \prod_{k} e^{-\frac{m}{2}|L_k|}\,\ga_{L_k} \\[4pt]
&\leq\, \sum_{n\geq2} \sum_{\,\substack{\T\!\text{ tree over}\\ \{1,\dots,n\}}} \frac{n}{n!} \sum_{(S_i)_{i=1}^n}\, \left(\prod_{i=1}^n e^{-\left(\beta\frac{\muh-\muv}{2}-\frac{m}{2}-\log4\right)|S_i|-\beta J}\right)\cdot \\ &\qquad\qquad \cdot e^{-\frac{m}{2}\dist_\h(S_1,x)}\,\ga_{S_1,x}\, \prod_{(i,j)\in\T}\!\!e^{-\frac{m}{2}\left(\dist_\h(S_i,S_j)-1\right)}\,\ga_{S_i,S_j} \ .
\end{split} \eeq
%
Substitute into \eqref{eq: CE II bound Sigma23} the entropy bound \eqref{eq: CE entropy treeA}:
\eq \label{eq: CE II boundentropy Sigma23} \begin{split}
\Sigma_2''' \,\leq\; &
\sum_{n\geq2} \sum_{\T\!\text{ tree over}\atop\{1,\dots,n\}} \frac{n}{n!} \sum_{(s_i)_{i=1,\dots,n}\atop s_i\text{ even }\geq2} \sum_{d_*\geq1} \sum_{(d_{ij})_{(i,j)\in\T}\atop d_{ij}\geq2} \left(\prod_{i=1}^n \frac{32}{3}\,4^{4s_i}\,s_i^{\deg_\T(i)}\right) \cdot\\
& \cdot\,\left(\prod_{i=1}^n e^{-\left(\beta\frac{\muh-\muv}{2}-\frac{m}{2}-\log4\right)s_i-\beta J}\right)\, e^{-\frac{m}{2}d_*}\,\bar\ga_{d_*}\, \prod_{(i,j)\in\T}\!\!e^{-\frac{m}{2}\left(d_{ij}-1\right)}\,\ga_{d_{ij}}
\end{split} \eeq
where $\ga_{d}:=\big( \frac{1}{2}e^{-\beta J}+e^{-\beta\frac{\muh+J}{2}\,(d-1)} \big)\,(1+o(1))$ and $\bar\ga_{d}:=\big( \frac{1}{\sqrt2}e^{-\beta\frac{J}{2}}+e^{-\beta\frac{\muh+J}{2}d} \big)\,(1+o(1))\,$. 
Observe that \eqref{eq: CE II boundentropy Sigma23} is identical to \eqref{eq: CE I boundentropy Sigma1} up to an extra factor $\sum_{d_*\geq1}e^{-\frac{m}{2}d_*}\,\bar\ga_{d_*}$, which equals
\[ \sum_{d_*\geq1}e^{-\frac{m}{2}d_*}\,\bar\ga_{d_*} \,=\,
\Big( \frac{1}{1-e^{-\frac{m}{2}}}\, \frac{e^{-\beta\frac{J}{2}}}{\sqrt2} \,+\, o(1) \Big)\, (1+o(1)) 
\,=\, \sqrt2\, e^{\beta\frac{\muh+2J}{2}}\, (1+o(1)) \]
since $1-e^{-\frac{m}{2}}=\frac{1}{2}\,e^{-\beta\frac{\muh+3J}{2}}\,(1+o(1))\,$.
Therefore we assume $\boldsymbol{\muh-\muv>\frac{\muh+J}{2}}$ and we exploit the inequality \eqref{eq: CE I boundfinal Sigma1} to bound the expression \eqref{eq: CE II boundentropy Sigma23}:
\eq \begin{split} \label{eq: CE II boundfinal Sigma23}
\Sigma_2''' \,&\leq\, \bigg(\frac{2^{26}e^2}{3}\bigg)^{\!2}\, e^{-\beta\,2(\muh-\muv)+\beta\frac{\muh+J}{2}}\; \sqrt2\, e^{\beta\frac{\muh+2J}{2}}\, (1+o(1)) \\ 
&=\, \frac{2^{52}e^4\sqrt2}{9}\; e^{-\beta\,2(\muh-\muv)+\beta\frac{2\muh+3J}{2}}\, (1+o(1)) \;.
\end{split} \eeq

Finally assume $n\geq2$ and that $L^x$ has one endpoint on $\partial_\v^\ext S_i$ and one on $\partial_\v^\ext S_j$ with $i\neq j\,$. By introducing an extra factor $n(n-1)/2$ we may assume that these endpoints lie on $\partial_\v^\ext S_1$ and $\partial_\v^\ext S_2$ respectively.
Fix a family of lines $(L_k)_{k=1}^p\in\PL_\La(\cup_iS_i)$ such that $\cup_kL_k\ni x$ (namely there exists $k\equiv k_x$ such that $L_k=L^x$), then consider the graph $G\equiv G\big((S_i)_i,(L_k)_k\big)$ with vertices $i\in\{1,\dots,n\}$ and edges $k\in\{1,\dots,p\}\,$: the edge $k$ joins the two vertices $i,j$ if the line $L_k$ has one endpoint on $\partial_\v^\ext S_i$ and the other one on $\partial_\v^\ext S_j\,$. $G$ admits at least one spanning sub-tree $\T$ that includes the edge $k_x\,$.
Therefore
\[\begin{split}
& \prod_{k=1}^p e^{-\frac{m}{2}|L_k|}\,\ga_{L_k} \,\leq\,
\prod_{k\in\T} e^{-\frac{m}{2}|L_k|}\,\ga_{L_k}  \,\leq\\[4pt]
& \leq\, e^{-\frac{m}{2}\left(\dist_\h(S_1,x)+\dist_\h(S_2,x)-1\right)}\,\bar\ga_{S_1,S_2,x}  \prod_{\substack{(i,j)\in\T\\ (i,j)\neq(1,2)}} e^{-\frac{m}{2}\left(\dist_\h(S_i,S_j)-1\right)}\,\ga_{S_i,S_j} 
\end{split}\]
where $\bar\ga_{S,S',x}:=\big( \frac{1}{2}e^{-\beta J}+e^{-\beta\frac{\muh+J}{2}\left(\dist_\h(S,x)+\dist_\h(S',x)-1\right)} \big)\,(1+o(1))$ and $\ga_{S,S'}:=\big( \frac{1}{2}e^{-\beta J}+e^{-\beta\frac{\muh+J}{2}\left(\dist_\h(S,S')-1\right)} \big)\,(1+o(1))\,$. 
Thus:
\eq \label{eq: CE II bound Lmax'} \begin{split}
&\max_{\substack{(L_k)_k,\,\cup_k\!L_k\ni x \\ L^x\text{ from }\partial_\v^\ext S_1\text{ to }\partial_\v^\ext S_2}}\,\prod_{k}e^{-\frac{m}{2}|L_k|}\,\ga_{L_k} \,\leq\\[4pt]
&\leq\, e^{-\frac{m}{2}\left(\dist_\h(S_1,x)+\dist_\h(S_2,x)-1\right)}\,\bar\ga_{S_1,S_2,x} \,\cdot\\ &\qquad\qquad\qquad\cdot \max_{\substack{\T\!\text{ tree over }\!\{1,\dots,n\}\\ \T\ni(1,2)}}\prod_{\substack{(i,j)\in\T\\ (i,j)\neq(1,2)}}\!\!\!\!\! e^{-\frac{m}{2}\left(\dist_\h(S_i,S_j)-1\right)}\,\ga_{S_i,S_j} \;.
\end{split} \eeq
Now using \eqref{eq: CE II bound Lsum} and \eqref{eq: CE II bound Lmax'} we can bound $\Sigma_2''''$:
\eq \label{eq: CE II bound Sigma24} \begin{split}
\Sigma_2'''' \,&:=\,
\sum_{n\geq2} \frac{1}{n!} \sum_{(S_i)_{i=1}^n} \left(\prod_{i} e^{-\left(\beta\frac{\muh-\muv}{2}-\frac{m}{2}\right)|S_i|-\beta J}\right) \cdot\\ &\qquad\qquad\qquad \cdot \sideset{}{^*}\sum_{\substack{(L_k)_k,\,\cup_k\!L_k\ni x\\ L^x\text{ from a }\partial_\v^\ext S_i\text{ to a }\partial_\v^\ext S_j\text{ with } i\neq j}} \prod_{k} e^{-\frac{m}{2}|L_k|}\,\ga_{L_k} \,\leq \\[4pt]
&\leq\, \sum_{n\geq2} \sum_{\substack{\,\T\!\text{ tree over }\!\{1,\dots,n\}\\ \T\ni(1,2)}}\!\!\!\! \frac{n(n-1)}{2\,n!} \sum_{(S_i)_{i=1}^n} \left(\prod_{i=1}^n e^{-\left(\beta\frac{\muh-\muv}{2}-\frac{m}{2}-\log4\right)|S_i|-\beta J}\right)\cdot \\ &\quad \cdot\, e^{-\frac{m}{2}\left(\dist_\h(S_1,x)+\dist_\h(S_2,x)-1\right)}\,\bar\ga_{S_1,S_2,x} \prod_{\substack{(i,j)\in\T\\ (i,j)\neq(1,2)}}\!\!\!\!\! e^{-\frac{m}{2}\left(\dist_\h(S_i,S_j)-1\right)}\,\ga_{S_i,S_j} \;.
\end{split} \eeq
Removing the edge $(1,2)$ from the tree $\T$ one obtains two disjoint trees $\T_1,\T_2$. By applying to each tree the entropy bound \eqref{eq: CE entropy treeA}, one finds:
\[\begin{split}
& \card\,\big\{ (S_i)_{i=1}^n\in\PS_\La \ \big|\; \dist_\h(S_1,x)=d_1\,,\ \dist_\h(S_2,x)=d_2\,,\ |S_i|=s_i\;\forall i\,,\\[2pt] &\qquad\qquad\qquad\qquad\qquad\qquad\qquad \dist_\h(S_i,S_j)=d_{ij}\;\forall(i,j)\in\T\!\smallsetminus\!(1,2) \big\} \,=\\[4pt]
& =\, \prod_{t=1,2} \card\,\big\{ (S_i)_{i\in\T_t}\in\PS_\La \ \big|\; \dist_\h(S_t,x)=d_t\,,\ |S_i|=s_i\;\forall i\in\T_t\,,\\[-10pt] &\qquad\qquad\qquad\qquad\qquad\qquad\qquad\qquad \dist_\h(S_i,S_j)=d_{ij}\;\forall(i,j)\in\T_t \big\} \,\leq \\[4pt]
&\leq\, \prod_{i=1}^n\left(\frac{32}{3}\,4^{4s_i}\,s_i^{\deg_{\T}(i)}\right) \;; 
\end{split}\]
then substitute this entropy bound into \eqref{eq: CE II bound Sigma24} and obtain:
\eq \label{eq: CE II boundentropy Sigma24} \begin{split}
\Sigma_2'''' \,\leq\; &
\sum_{n\geq2} \sum_{\substack{\,\T\!\text{ tree over }\!\{1,\dots,n\}\\ \T\ni(1,2)}}\!\!\!\! \frac{n(n-1)}{2\,n!} \sum_{\substack{(s_i)_{i=1,\dots,n}\\ s_i\text{ even }\geq2}} \sum_{\,d_1,d_2\geq1} \sum_{\substack{\;(d_{ij})_{(i,j)\in\T\smallsetminus(1,2)}\\ d_{ij}\geq2}} \\
&\ \left(\prod_{i=1}^n \frac{32}{3}\,4^{4s_i}\,s_i^{\deg_\T(i)}\right) 
\left(\prod_{i=1}^n e^{-\left(\beta\frac{\muh-\muv}{2}-\frac{m}{2}-\log4\right)s_i-\beta J}\right) \cdot\\
&\cdot e^{-\frac{m}{2}(d_1+d_2-1)}\,\ga_{d_1+d_2} \!\!\prod_{\substack{(i,j)\in\T\\ (i,j)\neq(1,2)}}\!\!\!\!e^{-\frac{m}{2}\left(d_{ij}-1\right)}\,\ga_{d_{ij}}
\end{split} \eeq
where $\ga_{d}:=\big( \frac{1}{2}e^{-\beta J}+e^{-\beta\frac{\muh+J}{2}\,(d-1)} \big)\,(1+o(1))\,$. 

\noindent As already seen, given $n\geq2$ and $\delta_1,\dots,\delta_n\geq1$, the number of trees $\T$ over the vertices $\{1,\dots,n\}\,$ with fixed degrees $\deg_\T(i)=\delta_i\,$ $\forall i=1,\dots,n$ is bounded by $\frac{(n-2)!}{(\delta_1-1)!\cdots(\delta_n-1)!}\,$.
Furthermore the number of edges of $\T$ different from $(1,2)$ is $n-2$.
Therefore the bound \eqref{eq: CE II boundentropy Sigma24} leads to:
\eq \label{eq: CE II boundrearr Sigma24} \begin{split}
\Sigma_2'''' \,\leq\,
\frac{1}{2}\, \sum_{n\geq2}\; & \Bigg( \frac{32}{3}\,e^{-\beta J}\, \sum_{s\geq2\atop\text{even}} e^{-\left(\beta\frac{\muh-\muv}{2}-\frac{m}{2}-5\log4\right)s}\, \sum_{\delta\geq1}\frac{s^{\delta}}{(\delta-1)!} \Bigg)^{\!n} \,\cdot\\
& \cdot \Bigg(\sum_{d\geq2}e^{-\frac{m}{2}\left(d-1\right)}\,\ga_{d}\Bigg)^{\!n-2} \cdot \sum_{d_1,d_2\geq1}\!\! e^{-\frac{m}{2}(d_1+d_2-1)}\,\ga_{d_1+d_2} \;.
\end{split} \eeq
The sums over $s,d$ have been already computed in \eqref{eq: CE I sums}, \eqref{eq: CE I sumd} respectively; the sum over $d_1,d_2$ gives, as $\beta\to\infty$,
\eq \label{eq: CE II sumd1d2} \begin{split}
\sum_{d_1,d_2\geq1}\!\! e^{-\frac{m}{2}(d_1+d_2-1)}\,\ga_{d_1+d_2} \,&=\,
\Bigg( \frac{1}{\big(1-e^{-\frac{m}{2}}\big)^2}\, \frac{e^{-\beta J}}{2} \,+\, o(1) \Bigg)\,(1+o(1)) \\
&=\, 2\,e^{\beta(\muh+2J)}\,(1+o(1)) \;.
\end{split} \eeq
Substitute \eqref{eq: CE I sums}, \eqref{eq: CE I sumd}, \eqref{eq: CE II sumd1d2} into \eqref{eq: CE II boundrearr Sigma24} and obtain
\eq \label{eq: CE II boundclose Sigma24}
\Sigma_2'''' \,\leq\,
\sum_{n\geq2}\; \Bigg( \frac{2^{26}e^2}{3}\,e^{-\beta(\muh-\muv)+\beta\frac{\muh+J}{2}}\,(1+o(1)) \Bigg)^{\!n}\; e^{\beta J}\,(1+o(1)) \;.
\eeq
Assume $\boldsymbol{\muh-\muv>\frac{\muh+J}{2}}\,$. Then for $\beta$ sufficiently large the \eqref{eq: CE II boundclose Sigma24} becomes:
\eq \label{eq: CE II boundfinal Sigma24} \begin{split}
\Sigma_2'''' \,&\leq\,
\left(\frac{2^{26}e^2}{3}\, e^{-\beta(\muh-\muv)+\beta\frac{\muh+J}{2}}\right)^2\, e^{\beta J}\,(1+o(1)) \\[2pt]
&=\, \frac{2^{52}e^4}{9}\; e^{-\beta\,2(\muh-\muv)+\beta(\muh+2J)}\, (1+o(1)) \;.
\end{split} \eeq

In conclusion, by using the estimates \eqref{eq: CE I n1}, \eqref{eq: CE I boundfinal Sigma1}, \eqref{eq: CE II n0}, \eqref{eq: CE II n1}, \eqref{eq: CE II boundfinal Sigma23}, \eqref{eq: CE II boundfinal Sigma24}, and the fact that $m=e^{-\beta\frac{\muh+3J}{2}}\,(1+o(1))$ (see lemma \ref{lem: 1D eigenval ratio}), if we assume $\boldsymbol{\muh-\muv>\frac{\muh+J}{2}}$, we find that:
\eq \label{eq: CE boundfinal} \begin{split}
& \frac{1}{m} \sum_{\substack{P\in\P_\La\\ \supp P\ni x}} \tilde\rho(P)\,e^{a(P)} \,=\\
&=\, e^{\beta\frac{\muh+3J}{2}}\, \big(\Sigma_1'+\Sigma_1''+\Sigma_2'+\Sigma_2''+\Sigma_2'''+\Sigma_2''''\big)\,(1+o(1)) \\
&\leq \Bigg( \frac{2^{24}}{3}\, e^{-\beta(\muh-\muv)+\beta\frac{\muh+J}{2}} \,+\,
\frac{2^{52}\,e^4}{9}\, e^{-\beta\,2(\muh-\muv)+\beta\frac{\muh+2J}{2}} \,+\,
\frac{1}{m}\,e^{-\frac{m}{2}N} \\
&\phantom{=\,\Bigg(} +\,\frac{2^{25.5}}{3}\, e^{-\beta(\muh-\muv)+\beta\frac{2\muh+5J}{2}} \,+\, \frac{2^{52.5}\,e^4}{9}\, e^{-\beta\,2(\muh-\muv)+\beta\frac{3\muh+6J}{2}} \\
&\phantom{=\,\Bigg(} +\, \frac{2^{52}\,e^4}{9}\, e^{-\beta\,2(\muh-\muv)+\beta\frac{3\muh+7J}{2}} \Bigg)\,(1+o(1)) \\
&=\, \Bigg( \frac{1}{m}\,e^{-\frac{m}{2}N} \,+\, \frac{2^{25.5}}{3}\, e^{\beta(\muv+\frac{5J}{2})} \Bigg)\,(1+o(1))
\end{split} \eeq
where $N$ is the minimum distance between two different vertical components of $\partial\La$ and $o(1)\to0$ as $\beta\to\infty$ (uniformly with respect to $N$).

\noindent Now we assume that $\boldsymbol{\muv+\frac{5J}{2}<0}$. Thus there exists $\beta_0>0$ such that for all $\beta>\beta_0$ the function $1+o(1)$ on the r.h.s. of \eqref{eq: CE boundfinal} is $<2$ and the term $\frac{2^{25.5}}{3}\, e^{\beta(\muv+\frac{5J}{2})} \leq 1/32\,$. There exists\footnote{$N_0=\frac{2}{m}\log\frac{32}{m}\,$.} also $N_0(\beta)$ such that for all $N>N_0(\beta)$ the term $\frac{1}{m}\,e^{-\frac{m}{2}N} \leq 1/32\,$.
Therefore if $\muv+\frac{5J}{2}<0$ (which entails also the previous condition $\muh-\muv>\frac{\muh+J}{2}$), then the inequality \eqref{eq: CE boundfinal} implies that
\[ \sum_{P\in\P_\La\atop\supp P\ni x}\!\! \tilde\rho(P)\,e^{a(P)} \,\leq\, \frac{m}{8} \]
for $\beta>\beta_0$ and $N>N_0(\beta)\,$. This concludes the proof. \qed
\endproof

\section{Proofs of the Liquid Crystal Properties} \label{sec: LiqCry}
In this section we will finally prove that the model behaves like a liquid crystal, as stated in the section \ref{sec: main}, by means of the cluster expansion results obtained in the previous sections.

\subsection{Proof of the theorem \ref{thm: LiqCry orientorder}}
We will prove the inequality \eqref{eq: LiqCry flrbound} for $f_{\l,x}\,$. That one for $f_{\r,x}\,$ can be proved analogously; then \eqref{eq: LiqCry fhbound} and \eqref{eq: LiqCry fl-frbound} follow since $f_{x}=f_{\l,x}+f_{\r,x}\,$.

\noindent Observe that
\[ \langle f_{\l,x} \rangle_\La^\h \,=\, \frac{Z_{\La\setminus x}^\h}{Z_\La^\h} \;,\]
where $Z_{\La\setminus x}^\h$ is the partition function over the lattice $\La\setminus x$ with  horizontal boundary conditions including a left-dimer at the site $x$.
Since $N>N_0(\beta)$ and $\dist_\h(x,\partial\La)>N_0(\beta)$, both partition functions satisfy the hypothesis of the corollary \ref{cor: CE}. Hence by the cluster expansion \eqref{eq: CE} the partition functions rewrite as
\[ Z_\La^\h \,=\, C_\La\, \exp\bigg(\; \sumPLa U_\La\big((P_t)_t\big) \bigg) \;,\]
\[ Z_{\La\setminus x}^\h \,=\, C_{\La\setminus x}\, \exp\bigg(\; \sumPLax U_{\La\setminus x}\big((P_t)_t\big) \bigg) \;.\]

\noindent By applying the definition \eqref{eq: Poly defC},
\[ \frac{C_{\La\setminus x}}{C_\La} \,=\,
\frac{b_{\r,x-(1,0)}\;b_{\l,x+(1,0)}}{\la_1} \;. \]

\noindent Now consider a polymer $P\in P_\La\cup\P_{\La\setminus x}\,$.
Keeping in mind the definitions of polymer \eqref{eq: Poly defP} and polymer activity \eqref{eq: Poly defrho}, 
observe that\footnote{The condition $\dist_\h(\supp P,x)>1$ guarantees that $\supp P\subseteq\La\setminus x$ and that the polymer $P$ does not include any line $L_k$ having one endpoint on $x\pm(1,0)$, nor any region $S_i$ containing these points.}
\[ \text{if }\dist_\h(\supp P,x)>1 \ \ \Rightarrow\ \ P\in\P_{\La}\cap\P_{\La\setminus x} \ ,\ \rho_{\La}(P)=\rho_{\La\setminus x}(P) \;.\]
Therefore:
\[\begin{split}
&\sumPLax U_{\La\setminus x}\big((P_t)_t\big) \;- \sumPLa U_{\La}\big((P_t)_t\big) \ \geq\\[2pt]
&\geq\ - \sumPLaxdist \big|U_{\La\setminus x}\big((P_t)_t\big)\big| \ - \sumPLadist \big|U_{\La}\big((P_t)_t\big)\big|  \ .
\end{split}\]
And by the inequalities \eqref{eq: CE clu-poly} and \eqref{eq: CE KPcond2} applied to both $Z_\La^\h$, $Z_{\La\setminus x}^\h\,$,
\[ \!\!\!\! \sumPLadist \big|U_{\La}\big((P_t)_t\big)\big| \ \leq
\sum_{\substack{P\in\P_\La\\ \dist_\h(\supp P,x)\leq1}}\!\!\!\!\!\! \widetilde\rho(P)\, e^{a(P)} \ \leq\;
3\,\frac{m}{8} \;;\]
\[ \!\!\!\! \sumPLaxdist \big|U_{\La\setminus x}\big((P_t)_t\big)\big| \ \leq
\sum_{\substack{P\in\P_{\La\setminus x}\\ \dist_\h(\supp P,x)\leq1}}\!\!\!\!\!\!\!\! \widetilde\rho(P)\, e^{a(P)} \ \leq\;
2\,\frac{m}{8} \;.\]

\noindent In conclusion one obtains:
\[\begin{split}
\langle f_{\l,x} \rangle_\La^\h \,=\, \frac{Z_{\La\setminus x}^\h}{Z_\La^\h} \,&\geq\,
\frac{b_{\r,\,x-(1,0)}\,b_{\l,\,x+(1,0)}}{\la_1}\; \exp\left(-5\,\frac{m}{8}\right) \\
&=\, \frac{1}{2}\,\big(1-e^{-\beta\frac{\muh+J}{2}}(1+o(1))\big) \;,
\end{split}\]
where the last identity follows from the fact that $\la_1\, b_{\r,x-(1,0)}\,b_{\l,x+(1,0)} \,=\, \LE^{(1)}B_{\r,x-(1,0)}\,B_{\l,x+(1,0)}\RE^{(1)} \,=\, \frac{1}{\sqrt2}(1-\frac{a}{2}\,(1+o(1)))\, \frac{1}{\sqrt2}(1-\frac{a}{2}\,(1+o(1)))$ (by lemma \ref{lem: 1D endpt La}, since there is a left-dimer fixed at $x$ according to $Z_{\La\setminus x}^\h$), $\la_1=1+\frac{ab}{2}\,(1+o(1))$ (proposition \ref{prop: 1D eigenval}), and $e^{-5m/8}=1-\frac{5}{8}ab\,(1+o(1))$ (lemma \ref{lem: 1D eigenval ratio}).
Finally, since $o(1)\to0$ as $\beta\to\infty$ and $o(1)$ does not depend on the choice of $x$ and $\La$, one may obtain the desired inequality eventually increasing $\beta_0\,$. \qed

\subsection{Proof of the corollary \ref{cor: LiqCry orientorder}}
Set $\varphi_{\La,N_0}:=\card\,\{x\in\La\,|\,\dist_\h(x,\partial\La)>N_0\}\,/\,|\La|\,$. By the theorem \ref{thm: LiqCry orientorder}, bound \eqref{eq: LiqCry fhbound}, using also $f_{\v,x}\leq1-f_{\h,x}$, one obtains:
\[ \langle \Deltao \rangle_\La^\h \,=\,
\frac{1}{|\La|}\,\sum_{x\in\La} \big( \langle f_{\h,x} \rangle_\La^\h - \langle f_{\v,x} \rangle_\La^\h \big) \,\geq\,
\varphi_{\La,N_0(\beta)}\, \big( 1-4\,e^{-\beta\frac{\muh+J}{2}} \big) \;.\]
On the other hand:
\[ \varphi_{\La,N_0} \,\geq\!
\min_{\substack{L\text{ maximal}\\ \text{ horiz.}\!\text{ line of }\La}} \varphi_{L,N_0} \,=\!
\min_{\substack{L\text{ maximal}\\ \text{ horiz.}\!\text{ line of }\La}}\! \frac{|L|-2N_0(\beta)}{|L|} \,=\,
1-2\,\frac{N_0(\beta)}{N} \;.\] \qed

\subsection{Proof of the corollary \ref{cor: LiqCry posorder1}}
Set $\varphi_{\La,N_0}:=\card\,\{x\in\La\,|\,\dist_\h(x,\partial\La)>N_0\}\,/\,|\La|\,$. By the theorem \ref{thm: LiqCry orientorder}, bound \eqref{eq: LiqCry fl-frbound},
\[ \big| \langle \Deltap \rangle_{\La}^{\h} \big| \,\leq\,
\frac{2}{|\La|} \sum_{x\in\La,\atop x_\h\!\text{ even}}\!\! \big|\langle f_{\r,x}\rangle_\La^\h - \langle f_{\l,x}\rangle_\La^\h\big| \,\leq\,
\varphi_{\La,N_0(\beta)}\,2e^{-\beta\frac{\muh+J}{2}} +\, 1-\varphi_{\La,N_0(\beta)} \;.\]
On the other hand we have already observed in the proof of the corollary \ref{cor: LiqCry orientorder} that $\varphi_{\La,N_0}\geq1-2N_0/N\,$. \qed

\subsection{Proof of the theorem \ref{thm: LiqCry corrdecay}}
We will prove the inequality \eqref{eq: LiqCry flfl}. \eqref{eq: LiqCry frfr} and \eqref{eq: LiqCry flfr} can be proved analogously.
First of all observe that, since $0\leq f_{\l,x},f_{\l,y}\leq1$, 
\eq \label{eq: LiqCry proofcorr diff-ratio}
\big| \langle f_{\l,x}\,f_{\l,y} \rangle_\La^\h - \langle f_{\l,x} \rangle_\La^\h\, \langle f_{\l,y} \rangle_\La^\h \big| \,\leq\, \log\left( \frac{\langle f_{\l,x}\,f_{\l,y} \rangle_\La^\h}{\langle f_{\l,x} \rangle_\La^\h\, \langle f_{\l,y} \rangle_\La^\h} \,\lor\, \frac{\langle f_{\l,x} \rangle_\La^\h\, \langle f_{\l,y} \rangle_\La^\h}{\langle f_{\l,x}\,f_{\l,y} \rangle_\La^\h} \right) \;.
\eeq
Now observe that:
\[ \langle f_{\l,x}\,f_{\l,y} \rangle_\La^\h \,=\, \frac{Z_{\La\setminus x,y}^\h}{Z_\La^\h} \ ,\quad
\langle f_{\l,x} \rangle_\La^\h \,=\, \frac{Z_{\La\setminus x}^\h}{Z_\La^\h} \ ,\quad
\langle f_{\l,y} \rangle_\La^\h \,=\, \frac{Z_{\La\setminus y}^\h}{Z_\La^\h} \ ,\]
where $Z_{\La\setminus x}^\h\,$, $Z_{\La\setminus y}^\h\,$, $Z_{\La\setminus x,y}^\h$ are the partition function respectively  over the lattices $\La\setminus x\,$, $\La\setminus y\,$, $\La\setminus x,y\,$, with horizontal boundary conditions including a left-dimer respectively at the site $x\,$, at the site $y\,$, at both sites $x,y\,$.
Therefore
\eq \label{eq: LiqCry proofcorr ratioZ}
\frac{\langle f_{\l,x}\,f_{\l,y} \rangle_\La^\h}{\langle f_{\l,x} \rangle_\La^\h\, \langle f_{\l,y} \rangle_\La^\h} \,=\,
\frac{Z_\La^\h\ Z_{\La\setminus x,y}^\h}{Z_{\La\setminus x}^\h\, Z_{\La\setminus y}^\h} \;.
\eeq
Since $N>N_0(\beta)\,$, $\dist_\h(x,\partial\La)>N_0(\beta)\,$, $\dist_\h(y,\partial\La)>N_0(\beta)\,$, $\dist_\h(x,y)>N_0(\beta)$, all four partition functions satisfy the hypothesis of the corollary \ref{cor: CE}. Hence by the cluster expansion \eqref{eq: CE} the partition functions rewrites as
\eq \label{eq: LiqCry proofcorr Zexpansion} \begin{split}
Z_\La^\h \,&=\, C_\La\, \exp\bigg(\; \sumPLa U_\La\big((P_t)_t\big) \bigg) \;, \\[2pt]
Z_{\La\setminus x}^\h \,&=\, C_{\La\setminus x}\, \exp\bigg(\; \sumPLax U_{\La\setminus x}\big((P_t)_t\big) \bigg) \;, \\[2pt]
Z_{\La\setminus y}^\h \,&=\, C_{\La\setminus y}\, \exp\bigg(\; \sumPLay U_{\La\setminus y}\big((P_t)_t\big) \bigg) \;, \\[2pt]
Z_{\La\setminus x,y}^\h \,&=\, C_{\La\setminus x,y}\, \exp\bigg(\; \sumPLaxy U_{\La\setminus x,y}\big((P_t)_t\big) \bigg) \;.
\end{split} \eeq

\noindent By applying the definition \eqref{eq: Poly defC}, it holds
\eq \label{eq: LiqCry proofcorr C}
\frac{C_\La\; C_{\La\setminus x,y}}{C_{\La\setminus x}\; C_{\La\setminus y}} \,=\, 1 \;.
\eeq

\noindent Now consider a polymer $P\in\P_\La\cup\P_{\La\setminus x}\cup\P_{\La\setminus y}\cup\P_{\La\setminus x,y}\,$. Keeping in mind the definitions of polymer \eqref{eq: Poly defP} and polymer activity \eqref{eq: Poly defrho}, observe that:
\[\begin{split}
& \text{if }\dist_\h(\supp P,x)>1 \;,\; \dist_\h(\supp P,y)>1 \ \ \Rightarrow\\[2pt]
& P\in\P_\La\cap\P_{\La\setminus x}\cap\P_{\La\setminus y}\cap\P_{\La\setminus x,y} \;,\; \rho_\La(P)=\rho_{\La\setminus x}(P)=\rho_{\La\setminus y}(P)=\rho_{\La\setminus x,y}(P) \;;
\end{split}\]
and that\footnote{The first possibility, namely $P$ polymer only of the lattices that contain $x$, happens when $\supp P\ni x$ or $P$ includes a region $S_i$ containing $x-(1,0)\,$. The second possibility, namely $P$ polymer only of the lattices that do not contain $x$, happens when $P$ includes a line $L_k$ with one endpoint on $x\pm(1,0)\,$. The last possibility happens when $P$ includes a region $S_i$ containing $x+(1,0)$ (and does not verify the other conditions).}:
\[\begin{split}
& \text{if }\dist_\h(\supp P,x)\leq1 \;,\; \dist_\h(\supp P,y)>1 \ \ \Rightarrow\\[2pt]
& P\in\big(\P_\La\cap\P_{\La\setminus y}\big)\setminus\big(\P_{\La\setminus x}\cup\P_{\La\setminus x,y}\big) \;,\; \rho_\La(P)=\rho_{\La\setminus y}(P) \;\text{ or}\\
& P\in\big(\P_{\La\setminus x}\cap\P_{\La\setminus x,y}\big)\setminus\big(\P_\La\cup\P_{\La\setminus y}\big) \;,\; \rho_{\La\setminus x}(P)=\rho_{\La\setminus x,y}(P) \;\text{ or}\\
& P\in\P_\La\cap\P_{\La\setminus x}\cap\P_{\La\setminus y}\cap\P_{\La\setminus x,y} \;,\; \rho_\La(P)=\rho_{\La\setminus y}(P) \;,\; \rho_{\La\setminus x}(P)=\rho_{\La\setminus x,y}(P) \;;
\end{split}\]
and the case $\dist_\h(\supp P,x)>1$, $\dist_\h(\supp P,y)\leq1$ is clearly symmetric to the previous one.
Therefore: 
\eq \label{eq: LiqCry proofcorr cluster} \begin{split}
& \sumPLa U_{\La}\big((P_t)_t\big) \ - \sumPLax U_{\La\setminus x}\big((P_t)_t\big) \ +\\[2pt]
&- \sumPLay U_{\La\setminus y}\big((P_t)_t\big) \ + \sumPLaxy U_{\La\setminus x,y}\big((P_t)_t\big) \ \leq\\[8pt]
&\leq \sumPLadistt \big|U_{\La}\big((P_t)_t\big)\big| \ + \sumPLaxdistt \big|U_{\La\setminus x}\big((P_t)_t\big)\big| \ +\\[2pt]
&\ \ + \sumPLaydistt \big|U_{\La\setminus y}\big((P_t)_t\big)\big| \ + \sumPLaxydistt \big|U_{\La\setminus x,y}\big((P_t)_t\big)\big| \ .
\end{split} \eeq
It is crucial to observe that given a cluster $(P_t)_t\in\CP_\La\,$, since $\cup_t\supp P_t$ have to be connected in $\Z^2\,$,
\[ \dist_{\Z^2}(x,y) \,\leq\, \dist_{\Z^2}(\cup_t\supp P_t,x) \,+\, \sum_{t}|\supp P_t|  -1 \,+\, \dist_{\Z^2}(\cup_t\supp P_t,y) \;.\]
Hence, assuming that $\dist_{\Z^2}(\cup_t\supp P_t,x)\leq1\,$, $\dist_{\Z^2}(\cup_t\supp P_t,y)\leq1\,$, it follows
\[ \begin{split}
&\prod_t \widetilde\rho(P_t) \,=\\
&=\, \prod_t\, \frac{1}{n_t!\,p_t!}\,\exp\left(-\beta\frac{\muh-\muv}{2}\,\sum_{i=1}^{n_t}|S_i|-m\,\sum_{k=1}^{p_t}\,|L_k|-\beta J\,n_t\right) \\[4pt]
&=\, \exp\left(-\frac{m}{4}\,\sum_t|\supp P_t|\right)\;\cdot\\
&\phantom{=\,} \cdot\, \prod_t\, \frac{1}{n_t!\,p_t!}\,\exp\left(-\bigg(\beta\frac{\muh-\muv}{2}-\frac{m}{4}\bigg)\,\sum_{i=1}^{n_t}|S_i|-\frac{3\,m}{4}\,\sum_{k=1}^{p_t}\,|L_k|-\beta J\,n_t\right) \\[4pt]
&\leq\, \exp\left(-\frac{m}{4}\,(\dist_{\Z^2}(x,y)-1)\right)\; \prod_t \widetilde\rho_*(P_t)
\end{split} \]
where $P_t=\big((S_i)_{i=1}^{n_t},(L_k)_{k=1}^{p_t}\big)$ for all $t$ and $\widetilde\rho_*(P_t)$ is defined as the factor appearing in the product over $t$ at the penultimate step.
By defining $a_*(P):=\frac{m}{4}|\supp P|\,$, we have that $\widetilde\rho_*(P)\,e^{a_*(P)}$ is essentially equivalent to $\widetilde\rho(P)\,e^{a(P)}\,$: we can follow exactly the proof of the theorem \ref{thm: CE KPcond} up to the inequality \eqref{eq: CE boundfinal} and prove that the Kotecky-Preiss conditions \eqref{eq: CE KPcond2}, \eqref{eq: CE KPcond} hold also with $\widetilde\rho_*\,$, $a_*$ and $m/16$ in place of $\widetilde\rho\,$, $a$ and $m/8$ (eventually increasing $\beta_0$).
Therefore, defining $\widetilde U_*\big((P_t)_t\big):= u\big((P_t)_t\big)\, \prod_t\widetilde\rho_*(P_t)\,$, by the general theory of cluster expansion the inequality \eqref{eq: CE clu-poly} holds also with $\widetilde U_*$, $\widetilde\rho_*$ and $a_*$ in place of $U_\La$, $\rho_\La$ and $a\,$.
As a consequence:
\eq \label{eq: LiqCry proofcorr cluster-dist} \begin{split}
& \sumPLadistt \big|U_{\La}\big((P_t)_t\big)\big| \ \leq
\sumPLadistt \big|u\big((P_t)_t\big)\big|\, \prod_t\widetilde\rho(P_t) \ \leq\\[4pt]
&\leq\; e^{-\frac{m}{4}(\dist_{\Z^2}(x,y)-1)} \sumPLadistt \big|u\big((P_t)_t\big)\big|\, \prod_t\widetilde\rho_*(P_t) \\[4pt]
&=\; e^{-\frac{m}{4}(\dist_{\Z^2}(x,y)-1)} \sumPLadistt \big|\widetilde U_*\big((P_t)_t\big)\big| \\[4pt] 
&\overset{\eqref{eq: CE clu-poly}}{\leq}\, e^{-\frac{m}{4}(\dist_{\Z^2}(x,y)-1)} \sum_{\substack{P\in\P_\La \\ \dist_\h(\supp P,\,x)\leq1}}\!\!\!\!\!\!\!\!\!\! \widetilde\rho_*(P)\,e^{a_*(P)} \\[4pt]
&\overset{\eqref{eq: CE KPcond2}}{\leq}\, e^{-\frac{m}{4}(\dist_{\Z^2}(x,y)-1)}\ \,3\,\frac{m}{16} \;.
\end{split} \eeq
The same reasoning can be repeated also for the clusters in $\CP_{\La\setminus x}$, $\CP_{\La\setminus y}$ and $\CP_{\La\setminus x,y}\,$. Thus, by \eqref{eq: LiqCry proofcorr ratioZ}, \eqref{eq: LiqCry proofcorr Zexpansion}, \eqref{eq: LiqCry proofcorr C}, \ref{eq: LiqCry proofcorr cluster}, \eqref{eq: LiqCry proofcorr cluster-dist}, one finally obtains:
\[ \frac{\langle f_{\l,x}\,f_{\l,y} \rangle_\La^\h}{\langle f_{\l,x} \rangle_\La^\h\, \langle f_{\l,y} \rangle_\La^\h} \,=\,
\frac{Z_\La^\h\ Z_{\La\setminus x,y}^\h}{Z_{\La\setminus x}^\h\, Z_{\La\setminus y}^\h} \,\leq\,
\exp\left( e^{-\frac{m}{4}(\dist_{\Z^2}(x,y)-1)}\;(3+2+2+2)\,\frac{m}{16} \right) \;.\]
The same bound can be shown to hold also for the inverse ratio $\frac{\langle f_{\l,x} \rangle_\La^\h\, \langle f_{\l,y} \rangle_\La^\h}{\langle f_{\l,x}\,f_{\l,y} \rangle_\La^\h} \,$, hence by \eqref{eq: LiqCry proofcorr diff-ratio} we conclude that:
\[ \big| \langle f_{\l,x}\,f_{\l,y} \rangle_\La^\h \,-\, \langle f_{\l,x} \rangle_\La^\h\, \langle f_{\l,y} \rangle_\La^\h \big| \,\leq\, e^{-\frac{m}{4}(\dist_{\Z^2}(x,y)-1)}\;\frac{9m}{16} \;.\] \qed

\subsection{Proof of the corollary \ref{cor: LiqCry posorder2}}
Since $\Deltap = \frac{2}{|\La|} \sum_{x\in\La,\atop x_\h\!\text{ even}} (f_{\r,x} - f_{\l,x})\,$, the variance of $\Delta$ rewrites as:
\[ \big\langle (\Deltap)^2 \big\rangle_{\La}^{\h} \,-\, \big(\langle \Deltap \rangle_{\La}^{\h}\big)^2 \,=\,
\frac{4}{|\La|^2} \sum_{x,y\in\La\atop x_\h,y_\h\!\text{ even}}\!\! C_{x,y} \]
with
\[\begin{split}
C_{x,y} \,:=\, & \left(\langle f_{\r,x}\,f_{\r,y} \rangle_\La^\h - \langle f_{\r,x} \rangle_\La^\h\, \langle f_{\r,y} \rangle_\La^\h\right) +
\left(\langle f_{\r,x} \rangle_\La^\h\, \langle f_{\l,y} \rangle_\La^\h - \langle f_{\r,x}\,f_{\l,y} \rangle_\La^\h\right) +\\
&+ \left(\langle f_{\l,x} \rangle_\La^\h\, \langle f_{\r,y} \rangle_\La^\h - \langle f_{\l,x}\,f_{\r,y} \rangle_\La^\h\right) +
\left(\langle f_{\l,x}\,f_{\l,y} \rangle_\La^\h - \langle f_{\l,x} \rangle_\La^\h\, \langle f_{\l,y} \rangle_\La^\h\right) \;.
\end{split}\]
By the theorem \ref{thm: LiqCry corrdecay}, for $x,y\in\La$ such that $\dist_\h(x,\partial\La)>N_0(\beta)$, $\dist_\h(y,\partial\La)>N_0(\beta)$ and $\dist_\h(x,y)>N_0(\beta)$, it holds
\[ C_{x,y} \,\leq\, 4\,\frac{9m}{16}\;e^{-\frac{m}{4}\,(\dist_{\Z^2}(x,y)-1)} \;.\]
Hence:
\[ \big\langle (\Deltap)^2 \big\rangle_{\La}^{\h} \,-\, \big(\langle \Deltap \rangle_{\La}^{\h}\big)^2 \,\leq\,
4\,\frac{9m}{16|\La|^2} \sum_{x,y\in\La\atop x\neq y} e^{-\frac{m}{4}\,(\dist_{\Z^2}(x,y)-1)} +\, 1-\varphi_{\La,\La,N_0(\beta)} \;,\]
where we set
\[ \varphi_{\La,\La'\!,N_0} := \frac{\card\,\{(x,y)\!\in\!\La\!\times\!\La'\,|\dist_\h(x,\partial\La)\lor\dist_\h(y,\partial\La')\lor\dist_\h(x,y)\!>\!N_0\}}{|\La|\;|\La'|} \;. \]
Now observe that
\[\begin{split}
\varphi_{\La,\La,N_0} \,&\geq\!
\min_{\substack{L,L'\text{ maximal}\\ \text{ horiz.}\!\text{ lines of }\La}} \varphi_{L,L',N_0} \,\geq\!
\min_{\substack{L,L'\text{ maximal}\\ \text{ horiz.}\!\text{ lines of }\La}}\!\! \frac{(|L|-2N_0)\,(|L'|-4N_0)}{|L|\,|L'|} \\[2pt] &\geq\, \left(1-2\frac{N_0}{N}\right) \left(1-4\frac{N_0}{N}\right) \;,
\end{split}\]
hence $1-\varphi_{\La,\La,N_0}\leq N_0/N\,(6-8N_0/N)\,$.
And on the other hand:
\[\begin{split}
&\sum_{x,y\in\La\atop x\neq y} e^{-\frac{m}{4}\,(\dist_{\Z^2}(x,y)-1)} \,\leq\,
|\La|\, \sum_{x\in\Z^2\atop x\neq0} e^{-\frac{m}{4}\,(\dist_{\Z^2}(x,0)-1)} \,=\\
&=\, |\La|\, \sum_{d\geq1} 4d\, e^{-\frac{m}{4}(d-1)} \,=\,
|\La|\, \frac{4}{(1-e^{-\frac{m}{4}})^2} \;. 
\end{split}\] \qed

\appendix

\section{Appendix: 1D Systems} \label{sec: 1D}
Consider a finite line $L$, that is a finite connected sub-lattice of $\Z$.
Consider a monomer-dimer model on $L$ given by the following partition function:
\[ Z_L \,=\, \sum_{\al\in\D_L} e^{-\beta H_L(\alpha)}\ e^{\Il(\al_{\xl})}\ e^{\Ir(\al_{\xr})} \;.\]
$\D_L$ denotes the set of monomer-dimer configurations on $L$ (allowing also external dimers at the endpoints of $L$);
the Hamiltonian is defined as
\[ H_L \,=\, \frac{\muh+J}{2}\;\card\left\{\parbox{6em}{\fn sites of $L$ with monomer}\right\} \,+\,
\frac{J}{2}\;\card\left\{\parbox{12em}{\fn sites of $L$ with dimer but neighbor to monomer in $L$}\right\} \;.\]
$\xl,\xr$ denote the left and the right endpoint of $L$ respectively; $\Il,\Ir$ represent the interaction among the configuration on $L$ and the boundary condition outside its endpoints.

This one-dimensional system can be described by a \textit{transfer matrix} $T$ over the three possible states of a site, $l\equiv$``left-dimer'', $r\equiv$``right-dimer'', $m\equiv$``monomer'':
\eq \label{eq: 1D defT}
T \,\equiv\, \bmat T(l,l) & T(l,r) & T(l,m) \\ T(r,l) & T(r,r) & T(r,m) \\ T(m,l) & T(m,r) & T(m,m) \emat \,:=\,
\bmat 0 & 1 & \sqrt{ab} \\ 1 & 0 & 0 \\ 0 & \sqrt{ab} & a \emat \;, \eeq
where to shorten the notation we set $\sqrt{a}:=e^{-\beta\frac{\muh+J}{4}}$ the transfer contribution of a monomer\footnote{The transfer energy of a monomer is half the energy of a monomer because it appears during two ``transfers''.}, $\sqrt{b}:=e^{-\beta\frac{J}{2}}$ the transfer contribution of a site with a dimer but neighbor to a monomer.
Two vectors are also needed to encode the boundary conditions:
\eq \begin{split} \label{eq: 1D defB}
\LB \,\equiv\, \bmat \LB(l) & \LB(r) & \LB(m) \emat \,:=\, \bmat e^{\Il(l)} & e^{\Il(r)} & \sqrt{a}\,e^{\Il(m)} \emat \;,\\[4pt]
\RB \,\equiv\, \bmat \RB(l) \\ \RB(r) \\ \RB(m) \emat \,:=\, \bmat e^{\Ir(l)} \\ e^{\Ir(r)} \\ \sqrt{a}\,e^{\Ir(m)} \emat \;. \qquad\qquad\quad
\end{split} \eeq

\begin{proposition} \label{prop: 1D Zbilin}
The partition function of the system rewrites as a bilinear form:
\eq \label{eq: 1D Zbilin}
Z_L \,=\, \LB\;T^{|L|-1}\,\RB \;. \eeq
\end{proposition}

\proof
According to the previous definitions it is clear that for every configuration $\al\in\{l,r,m\}^{|L|}$
\begin{multline*}
\1(\al\in\D_L)\ e^{-\beta H_L(\al)} \,=\\
=\, \sqrt{a}^{\,\1(\al_1=m)}\; T(\al_1,\al_2)\;T(\al_2,\al_3)\,\dots\,T(\al_{|L|-1},\al_{|L|})\; \sqrt{a}^{\,\1(\al_{|L|}=m)}\ \;.
\end{multline*}
Therefore
\[\begin{split}
Z_L \,&= \!\sum_{\,\al\in\{l,r,m\}^{|L|}}\!\! \LB(\al_1)\; T(\al_1,\al_2)\;T(\al_2,\al_3)\,\dots\,T(\al_{|L|-1},\al_{|L|})\; \RB(\al_{|L|}) \\
&=\, \LB\;T^{|L|-1}\,\RB \;.
\end{split}\] \qed
\endproof

Assume for the moment that the transfer matrix $T$ is diagonalizable. Denote by $\la_1,\la_2,\la_3$ its eigenvalues and by $\RE^{(1)},\RE^{(2)},\RE^{(3)}$, $\LE^{(1)},\LE^{(2)},\LE^{(3)}$ the corresponding right (column) eigenvectors and left (row) eigenvectors, normalized so that $\LE^{(i)} \RE^{(i)} = 1$ for $i=1,2,3$.

\begin{corollary} \label{cor: 1D Zeigen}
\eq \label{eq: 1D Zeigen}
Z_L \,=\, \sum_{i=1,2,3} \la_i^{|L|-1}\, \LB\,\RE^{(i)}\,\LE^{(i)}\,\RB \;. \eeq
\end{corollary}

\proof
Since we are assuming that $T$ is diagonalizable, it holds $T=P\,D\,P^{-1}$ where $D$ is the diagonal matrix of eigenvalues, $P$ is the matrix with the right eigenvectors on the columns, $P^{-1}$ has the left eigenvectors on the rows.
Then $T^{|L|-1}=P\,D^{|L|-1}\,P^{-1}$ and
\[ \LB\;T^{|L|-1}\,\RB\,=\, (\LB\,P)\,D^{|L|-1}\,(P^{-1}\,\RB) \,=\, \sum_{i=1}^3 \,(\LB\,\RE^{(i)})\,\la_i^{|L|-1}\,(\LE^{(i)}\RB) \;. \] \qed
\endproof

Now our purpose is to diagonalise the transfer matrix $T$ when $\beta$ is large.

\begin{proposition} \label{prop: 1D eigenval}
For all $\beta>0$ the transfer matrix $T$ is diagonalizable over $\R$.
Its eigenvalues are
\eq \label{eq: 1D eigenval} \begin{split} 
& \la_1 \,=\, 1 + \frac{ab}{2}\, (1+o(1)) \\
& \la_2 \,=\, -1 + \frac{ab}{2}\, (1+o(1)) \\
& \la_3 \,=\, a - ab- a^3b\, (1+o(1))
\end{split} \eeq
as $\beta\to\infty\,$. 
\end{proposition}

\proof
The eigenvalues $\la_1,\la_2,\la_3$ are the (complex) roots of the characteristic polynomial of $T$, that is
\[ p(\la) := \det(\la I-T) = -ab + (\la-a)(\la^2-1) \;.\]
For all $\beta>0$ it turns out that $p$ has 3 distinct real roots%
\footnote{The discriminant of the cubic is $\Delta=18a(1\!-\!b)+4a^2(1\!-\!b)+a^2+4-27a^2(1\!-\!b)$, which is strictly positive for all $0\leq a,b\leq1$, $(a,b)\neq(1,0)$.}%
, hence $T$ is diagonalizable over the reals.\\
As $\beta\to\infty$, $p(\la)\to\la(\la^2-1)$ hence $\la_1\to1\,$, $\la_2\to-1\,$, $\la_3\to0\,$. 
Thus it is convenient to write $\la_1=1+\eps_1\,$, $\la_2=-1+\eps_2\,$, $\la_3=a+\eps_3\,$ with $\eps_i\to0$ as $\beta\to\infty$ for $i=1,2,3$.
Now expand the polynomial $p$ in powers of $\eps_i$ and truncate it at the first order:
\[\begin{split}
& 0 \,=\, p(\la_1) \,=\, -ab+(1-a+\eps_1)\,(2\eps_1+\eps_1^2) \,=\, -ab+2\eps_1\,(1+o(1)) \\
&\quad \Rightarrow\, \eps_1=\frac{ab}{2}\,(1+o(1)) \;;\\[2pt]
& 0 \,=\, p(\la_2) \,=\, -ab+(-1-a+\eps)\,(-2\eps_2+\eps_2^2) \,=\, -ab+2\eps_2\,(1+o(1)) \\
&\quad \Rightarrow\, \eps_2=\frac{ab}{2}\,(1+o(1)) \;;\\[2pt]
& 0 \,=\, p(\la_3) \,=\, -ab+\eps_3\,\big((a+\eps_3)^2-1\big) \,=\, -ab-\eps_3\,(1+o(1)) \\
&\quad \Rightarrow\, \eps_3=-ab\,(1+o(1)) \;.
\end{split}\]
In order to find the following order of $\la_3$, now one can write $\la_3=a-ab\,(1+\eps_3')$ with $\eps_3'\to0$ as $\beta\to\infty$ and repeat the procedure:
\[\begin{split}
& 0 \,=\, \frac{p(\la_3)}{-ab} \,=\, 1 + (1+\eps_3')\left(a^2\,(1+o(1))-1\right) \,=\, a^2\,(1+o(1)) - \eps_3'\,(1+o(1)) \\
&\quad \Rightarrow\, \eps_3'=a^2\,(1+o(1)) \;. 
\end{split}\] \qed
\endproof

\begin{proposition} \label{prop: 1D eigenvec}
The right eigenvectors of the transfer matrix $T$ are
\eq \label{eq: 1D eigenvec} \begin{split}
& \RE^{(1)} \,=\, \frac{1}{\sqrt2}\, \bmat 1-\frac{a}{2}\,(1+o(1)) \\ 1-\frac{a}{2}\,(1+o(1)) \\ \sqrt{ab}\,(1+o(1)) \emat \\[2pt]
& \RE^{(2)} \,=\, \frac{1}{\sqrt2}\, \bmat 1+\frac{a}{2}\,(1+o(1)) \\ -1-\frac{a}{2}\,(1+o(1)) \\ \sqrt{ab}\,(1+o(1)) \emat \\[2pt]
& \RE^{(3)} \,=\, \bmat -a\sqrt{ab}\,(1+o(1)) \\ -\sqrt{ab}\,(1+o(1)) \\ 1+a\,(1+o(1)) \emat
\end{split} \eeq
and moreover
\[\begin{split}
& \RE^{(2)}(1)+\RE^{(2)}(2)+\sqrt{ab}\,\RE^{(2)}(3) \,=\, \frac{ab}{2\sqrt2}\,(1+o(1)) \\
& \RE^{(3)}(2)+\sqrt{ab}\,\RE^{(3)}(3) \,=\, -a^2\sqrt{ab}\,(1+o(1))
\end{split}\]
as $\beta\to\infty$.
The left eigenvectors are obtained by a simple transformation: $\LE^{(i)} = \sigma\big(\RE^{(i)}\big)\,$ for $i=1,2,3$, where
\[ \sigma \bmat v_1 \\ v_2 \\ v_3 \emat \,:=\, \bmat v_2 & v_1 & v_3 \emat \;.\]
\end{proposition}

\proof
The right eigenvectors $\RE$ associated to the eigenvalue $\la$ are the non-zero solutions of the linear system
\[ (\la I - T)\,\RE = 0 \ \Leftrightarrow\ \RE = \bmat \la(\la-a) \\ \la-a \\ \sqrt{ab} \emat t \ ,\ t\in\R \;.\]
And the left eigenvectors $\LE$ associated to the same eigenvalue $\la$ are the non-zero solutions of the linear system
\[ \LE\,(\la I - T) = 0 \ \Leftrightarrow\ \LE= \bmat \la-a & \la(\la-a) & \sqrt{ab}\, \emat t \ ,\ t\in\R \;.\]
The desired normalization $\LE\,\RE = 1$ can be obtained by choosing
\[ t=\sqrt{2\la(\la-a)+ab} \]
in both cases.
Now to conclude the proof it is sufficient to exploit the estimates of the eigenvalues given by the proposition \ref{prop: 1D eigenval}.
\qed
\endproof

The formula \eqref{eq: 1D Zeigen} together with the estimates of propositions \ref{prop: 1D eigenval}, \ref{prop: 1D eigenvec} give us a complete control on the one-dimensional system on $L$ at low temperature, for every choice of the boundary conditions.

We concentrate on providing an estimation of the quantity $R_L$ defined by \eqref{eq: Poly defRL}, since it is needed in the section \ref{sec: Poly}.
We have to distinguish three cases, according to where the endpoints of $L$ lie.

\begin{lemma} \label{lem: 1D eigenval ratio}
The ratios of the eigenvalues of the transfer matrix $T$ are
\[ \frac{\la_2}{\la_1} \,=\, -1+ab\,(1+o(1)) \quad,\quad \frac{\la_3}{\la_2} \,=\, -a+ab\,(1+o(1)) \]
as $\beta\to\infty$.
In particular setting $m:=-\log\big|\la_2/\la_1\big|$ it holds
\eq \label{eq: 1D m}
e^{-m} \,=\, 1-e^{-\beta\frac{\muh+3J}{2}}\,(1+o(1)) \quad\text{as }\beta\to\infty \;. \eeq
\end{lemma}

\proof
It follows immediately from the proposition \ref{prop: 1D eigenval}. 
\qed
\endproof

\begin{lemma} \label{lem: 1D endpt S}
If $\xl\in\partial^\ext_\r S_j$, then as $\beta\to\infty$
\[ \begin{split}
& \LB\,\RE^{(1)} = \frac{\sqrt{b}}{\sqrt{2}}\,(1+o(1)) \;\\
& \LB\,\RE^{(2)} = -\frac{\sqrt{b}}{\sqrt{2}}\,(1+o(1)) \;\\ 
& \LB\,\RE^{(3)} = \sqrt{a}\,(1+o(1)) \;.
\end{split} \]
If $\xr\in\partial^\ext_\l S_j$, then the same estimates hold for $\LE^{(1)}\RB\,$, $\LE^{(2)}\RB\,$, $\LE^{(3)}\RB$ respectively.
\end{lemma}

\proof
If $\xl\in\partial^\ext_\r S_j$ then by \eqref{eq: Poly BCendptl} and \eqref{eq: 1D defB} the vector describing the boundary condition on the left side of the line $L$ is $\LB = \bmat 0 & \sqrt{b} & \sqrt{a} \emat\,$.
Then the estimates for $\LB\,\RE^{(i)}$, $i=1,2,3$, are computed using the proposition \ref{prop: 1D eigenvec}. \qed
\endproof

\begin{lemma} \label{lem: 1D endpt La}
If $\xl\in\partial_\l\La$, then as $\beta\to\infty$
\[ \LB\,\RE^{(1)} = \begin{cases}\ \frac{1}{\sqrt{2}}\,\big(1-\frac{a}{2}\,(1+o(1))\big) & \text{if the h-dimer on $\xl\!-\!(1,0)$ has fixed position} \\[2pt]\ \sqrt{2}\,\big(1-\frac{a}{2}\,(1+o(1))\big) & \text{if the h-dimer on $\xl\!-\!(1,0)$ has free position} \end{cases} \]
\[ \LB\,\RE^{(2)} = \begin{cases}\ -\frac{1}{\sqrt{2}}\,\big(1+\frac{a}{2}\,(1+o(1))\big) & \text{if the h-dimer on $\xl\!-\!(1,0)$ is fixed to the left} \\[2pt]\ \frac{1}{\sqrt{2}}\,\big(1+\frac{a}{2}\,(1+o(1))\big) & \text{if the h-dimer on $\xl\!-\!(1,0)$ is fixed to the right} \\[2pt]\ \frac{ab}{2\sqrt{2}}\,(1+o(1)) & \text{if the h-dimer on $\xl\!-\!(1,0)$ has free position} \end{cases} \]
\[ \LB\,\RE^{(3)} = \begin{cases}\ -a^2\sqrt{ab}\,(1+o(1)) & \text{if the h-dimer on $\xl\!-\!(1,0)$ is fixed to the left} \\[2pt]\  -a\sqrt{ab}\,(1+o(1)) & \text{if the h-dimer on $\xl\!-\!(1,0)$ is fixed to the right or free} \end{cases} \]
If $\xr\in\partial_\r\La$, then the same estimates hold respectively for $\LE^{(1)}\RB\,$, $\LE^{(2)}\RB\,$, $\LE^{(3)}\RB$ after substituting: $\xl-(1,0)$ by $\xr+(1,0)\,$, ``left'' by ``right'' and ``right'' by ``left''.
\end{lemma}

\proof
If $\xl\in\partial_\l\La$ then by \eqref{eq: Poly BCendptl} and \eqref{eq: 1D defB} the vector describing the boundary condition on the left side of the line $L$ is: $\LB = \bmat 0 & 1 & \sqrt{ab} \emat$ if a left-dimer is fixed on $\xl\!-\!(1,0)$; $\LB = \bmat 1 & 0 & 0 \emat$ if a right-dimer is fixed on $\xl\!-\!(1,0)$; $\LB = \bmat 1 & 1 & \sqrt{ab} \emat$ if on $\xl\!-\!(1,0)\,$ there is a h-dimer with free position.
Then the estimates for $\LB\,\RE^{(i)}$, $i=1,2,3$, are computed using the proposition \ref{prop: 1D eigenvec}. \qed
\endproof

\begin{lemma} \label{lem: 1D endpt La}
If $\xl\in\partial_\l\La$, then as $\beta\to\infty$
\[ \LB\,\RE^{(1)} = \begin{cases}\ \frac{1}{\sqrt{2}}\,\big(1-\frac{a}{2}\,(1+o(1))\big) & \text{if the h-dimer on $\xl\!-\!(1,0)$ has fixed position} \\[2pt]\ \sqrt{2}\,\big(1-\frac{a}{2}\,(1+o(1))\big) & \text{if the h-dimer on $\xl\!-\!(1,0)$ has free position} \end{cases} \]
\[ \LB\,\RE^{(2)} = \begin{cases}\ -\frac{1}{\sqrt{2}}\,\big(1+\frac{a}{2}\,(1+o(1))\big) & \text{if the h-dimer on $\xl\!-\!(1,0)$ is fixed to the left} \\[2pt]\ \frac{1}{\sqrt{2}}\,\big(1+\frac{a}{2}\,(1+o(1))\big) & \text{if the h-dimer on $\xl\!-\!(1,0)$ is fixed to the right} \\[2pt]\ \frac{ab}{2\sqrt{2}}\,(1+o(1)) & \text{if the h-dimer on $\xl\!-\!(1,0)$ has free position} \end{cases} \]
\[ \LB\,\RE^{(3)} = \begin{cases}\ -a^2\sqrt{ab}\,(1+o(1)) & \text{if the h-dimer on $\xl\!-\!(1,0)$ is fixed to the left} \\[2pt]\  -a\sqrt{ab}\,(1+o(1)) & \text{if the h-dimer on $\xl\!-\!(1,0)$ is fixed to the right or free} \end{cases} \]
If $\xr\in\partial_\r\La$, then the same estimates hold respectively for $\LE^{(1)}\RB\,$, $\LE^{(2)}\RB\,$, $\LE^{(3)}\RB$ after substituting: $\xl-(1,0)$ by $\xr+(1,0)\,$, ``left'' by ``right'' and ``right'' by ``left''.
\end{lemma}

\proof
If $\xl\in\partial_\l\La$ then by \eqref{eq: Poly BCendptl} and \eqref{eq: 1D defB} the vector describing the boundary condition on the left side of the line $L$ is: $\LB = \bmat 0 & 1 & \sqrt{ab} \emat$ if a left-dimer is fixed on $\xl\!-\!(1,0)$; $\LB = \bmat 1 & 0 & 0 \emat$ if a right-dimer is fixed on $\xl\!-\!(1,0)$; $\LB = \bmat 1 & 1 & \sqrt{ab} \emat$ if on $\xl\!-\!(1,0)\,$ there is a h-dimer with free position.
Then the estimates for $\LB\,\RE^{(i)}$, $i=1,2,3$, are computed using the proposition \ref{prop: 1D eigenvec}. \qed
\endproof

\begin{proposition} \label{prop: 1D RL bounds}
Denote by $o(1)$ any function $\omega(\beta,\muh,J)$ that goes to zero as $\beta\to\infty$ and does not depend on the choice of the line $L$ nor on $\La$. Then for every line $L\in\L_\La(\cup_jS_j)$, $S_j\in\S_\La$ pairwise disconnected, $\La\subset\Z^2$ finite, it holds
\eq \label{eq: 1D bound R}
|R_L| \,\leq\, e^{-m|L|}\,\ga_L
\eeq
where the quantity $\ga_L$ can be chosen as follows:
\eq \label{eq: Poly gamma}
\ga_{L} := \begin{cases} \left( \frac{e^{-\beta J}}{2} + e^{-\beta\frac{\muh+J}{2}|L|} \right) (1+o(1)) & \text{if }\xl\in\cup_i\partial_\r^\ext S_i\,,\ \xr\in\cup_i\partial_\l^\ext S_i \\[4pt]
\frac{e^{-\beta\frac{J}{2}}}{\sqrt{2}}\,(1+o(1)) & \text{if }\xl\in\cup_i\partial_\r^\ext S_i\,,\ \xr\in\cup_i\partial_\r\La \\ & \text{or vice versa }\xl\in\cup_i\partial_\l\La\,,\ \xr\in\cup_i\partial_\l^\ext S_i \\[2pt]
1+o(1) & \text{if }\xl\in\partial_\l\La\,,\ \xr\in\partial_\r\La \end{cases}
\eeq
\end{proposition}

\proof
$\bullet$ Suppose $\xl\in\partial^\ext_\r S_i$ and $\xr\in\partial^\ext_\l S_j\,$. The definition \eqref{eq: Poly defRL} and the corollary \ref{cor: 1D Zeigen} give
\[\begin{split}
\la_1\,R_L \,&=\, \frac{Z_L}{\la_1^{|L|-1}} \,-\, \LB\RE^{(1)}\LE^{(1)}\RB \\
&=\, \left(\frac{\la_2}{\la_1}\right)^{\!|L|-1}\!\LB\RE^{(2)}\LE^{(2)}\RB \,+\, \left(\frac{\la_3}{\la_1}\right)^{\!|L|-1}\!\LB\RE^{(3)}\LE^{(3)}\RB \;.
\end{split}\]
By the lemma \ref{lem: 1D eigenval ratio} $|\la_3/\la_1|\leq a\,|\la_2/\la_1|$ when $\beta$ is sufficiently large. Therefore, using also the estimates of lemma \ref{lem: 1D endpt S}, one finds
\[\begin{split}
|R_L| \,&\leq\, \left|\frac{\la_2}{\la_1}\right|^{|L|-1} \left(\frac{b}{2}\,(1+o(1)) \,+\, a^{|L|-1}\,a\,(1+o(1))\right) \\
&=\, \left|\frac{\la_2}{\la_1}\right|^{|L|-1} \left( \frac{b}{2} + a^{|L|} \right)(1+o(1)) \;.
\end{split}\]
%
$\bullet$ Suppose now $\xl\in\partial^\ext_\r S_j$ and $\xr\in\partial_\r\La\,$. The definition \eqref{eq: Poly defRL} and the corollary \ref{cor: 1D Zeigen} give
\[\begin{split}
\la_1^{1/2}\,R_L \,&=\, \frac{Z_L}{\la_1^{|L|-1}\LE^{(1)}\RB} \,-\, \LB\RE^{(1)}\\[2pt]
&=\,  \left(\frac{\la_2}{\la_1}\right)^{\!|L|-1} \frac{\LB\RE^{(2)}\LE^{(2)}\RB}{\LE^{(1)}\RB} \,+\, \left(\frac{\la_3}{\la_1}\right)^{\!|L|-1} \frac{\LB\RE^{(3)}\LE^{(3)}\RB}{\LE^{(1)}\RB} \;.
\end{split}\]
By the lemma \ref{lem: 1D eigenval ratio} $|\la_3/\la_1|\leq a\,|\la_2/\la_1|$ when $\beta$ is sufficiently large. Therefore, using also the estimates of lemmas \ref{lem: 1D endpt S}, \ref{lem: 1D endpt La}, one obtains
\[ |R_L| \,\leq\, \left|\frac{\la_2}{\la_1}\right|^{|L|-1} \gamma \]
with
\[\begin{split}
\gamma \,&= \begin{cases}\text{if fixed h-dimer on $\xr\!+\!(1,0)$} & \frac{\sqrt{b}}{\sqrt{2}}\,(1+o(1)) \,+\, a^{|L|-1}\,O(a^2\sqrt{b}\,) \\[2pt]
\text{if free h-dimer on $\xr\!+\!(1,0)$} & \frac{ab\sqrt{b}}{4\sqrt2}\,(1+o(1)) \,+\, a^{|L|-1}\,\frac{a^2b}{\sqrt2}\,(1+o(1))
\end{cases}\\
&= \begin{cases} \ \frac{\sqrt{b}}{\sqrt2}\,(1+o(1)) \\[2pt] \ \left( \frac{ab\sqrt{b}}{4\sqrt2}+\frac{a^{|L|+1}b}{\sqrt2} \right) (1+o(1)) \end{cases} 
\!\leq\; \frac{\sqrt{b}}{\sqrt2}\,(1+o(1)) \;.
\end{split}\]
%
$\bullet$ Suppose now $\xl\in\partial_\l\La$ and $\xr\in\partial_\r\La\,$. The definition \eqref{eq: Poly defRL} and the corollary \ref{cor: 1D Zeigen} give
\[\begin{split}
R_L \,&=\, \frac{Z_L}{\la_1^{|L|-1}\LB\RE^{(1)}\LE^{(1)}\RB} \,-\, 1 \\[2pt]
&=\,  \left(\frac{\la_2}{\la_1}\right)^{\!|L|-1} \frac{\LB\RE^{(2)}\LE^{(2)}\RB}{\LB\RE^{(1)}\LE^{(1)}\RB} \,+\, \left(\frac{\la_3}{\la_1}\right)^{\!|L|-1} \frac{\LB\RE^{(3)}\LE^{(3)}\RB}{\LB\RE^{(1)}\LE^{(1)}\RB}  \;.
\end{split}\]
By the lemma \ref{lem: 1D eigenval ratio} $|\la_3/\la_1|\leq a\,|\la_2/\la_1|$ when $\beta$ is sufficiently large. Therefore, using also the estimates of lemma \ref{lem: 1D endpt La}, one obtains
\[ |R_L| \,\leq\, \left|\frac{\la_2}{\la_1}\right|^{|L|-1} \gamma \]
with
\[\begin{split}
\gamma \,&= \begin{cases}\text{if fixed h-b.c. on both sides} & 1+2a\,(1+o(1)) \,+\, a^{|L|-1}\, O(a^3b) \\[2pt]
\text{if fixed h-b.c. on one side,}  & \frac{ab}{4}\,(1+o(1)) \,+\, a^{|L|-1}\,O(a^3b) \\[-4pt] \text{free h-b.c. on the other one} & \\[2pt]
\text{if free h-b.c. on both sides} & \frac{a^2b^2}{8}\,(1+o(1)) \,+\, a^{|L|-1}\,\frac{a^3b}{2}\,(1+o(1)) \end{cases}\\
&= \begin{cases} \ 1+2a\,(1+o(1)) \\[2pt] \ \frac{ab}{4}\,(1+o(1)) \\[2pt] \ \left( \frac{a^2b^2}{8}+\frac{a^{|L|+2}b}{2} \right) (1+o(1))
\end{cases}
\!\leq\; 1+o(1) \;.
\end{split}\]
\qed
\endproof

\section{Appendix: Cluster Expansion} \label{sec: absCE}
In this Appendix we state the main results about the general theory of cluster expansion used in this paper. The condition that we adopt to guarantee the convergence of the expansion is due to Kotecky-Preiss \cite{KP}. For a modern proof we refer to \cite{U}.

Let $\P$ be a finite set, called the \textit{set of polymers}.
Let $\rho:\P\to\C$, called the \textit{polymer activity}, and $\delta:\P\times\P\to\{0,1\}$, called the \textit{polymer hard-core interaction}, such that $\delta(P,P)=0\,$ and $\delta(P,P')=\delta(P',P)$ for all $P,P'\in\P$.
Consider the \textit{polymer partition function}:
\eq \label{eq: absCE Z} \begin{split}
\ZZ \,&:=\, \sum_{\P'\subseteq\P}\; \prod_{P\in\P'}\rho(P) \prod_{\substack{P,P'\in\P'\\ P\neq P'}}\!\!\!\!\delta(P,P') \\
&=\, \sum_{q\geq0}\,\frac{1}{q!}\sum_{P_1,\dots,P_q\in\P}\, \prod_{t=1}^q\rho(P_t)\; \prod_{t<s}\delta(P_t,P_s) \;.
\end{split} \eeq
A family of polymers $(P_1,\dots,P_q)$ is called \textit{compatible} if $\delta(P_t,P_s)=1$ for all $t\neq s\,$; otherwise it is called \textit{incompatible}. Observe that in the partition function $\ZZ$ only the compatible families of polymers give non-zero contributions.\\
A family of polymers $(P_1,\dots,P_q)$ is called a \textit{cluster} if the graph with vertex set $\{1,\dots,q\}$ and edge set $\{(t,s)\,|\,\delta(P_t,P_s)=0\}$ is connected.

\begin{theorem} \label{thm: absCE main}
Suppose that there exists $a\!:\P\to[0,\infty[\,$, called size function, such that the Kotecky-Preiss condition is satisfied, namely:
\eq \label{eq: absCE KPcondition}
\sum_{\substack{P\in\P\\ \delta(P,P^*)=0}} |\rho(P)|\,e^{a(P)} \,\leq\, a(P^*) \quad\forall\,P^*\!\in\P \;. \eeq
Then:
\eq \label{eq: absCE logZ}
\log\ZZ \,=\, \sum_{q\geq0}\,\frac{1}{q!}\sum_{P_1,\dots,P_q\in\P} \left(\prod_{t=1}^q\rho(P_t)\right) u(P_1,\dots,P_q) \eeq
where the series on the r.h.s. is absolutely convergent and
\eq \label{eq: absCE u}
u(P_1,\dots,P_q) \,:= \sum_{\substack{G=(V,E)\!\text{ connected graph}\\ V=\{1,\dots,q\}\\ E\subseteq\{(t,s)\,|\,\delta(P_t,P_s)=0\}}} (-1)^{|E|} \;.\eeq
Moreover, for all $\mathscr{E}\subseteq\P$
\eq \label{eq: absCE event}
\sum_{q\geq0}\,\frac{1}{q!}\sum_{\substack{P_1,\dots,P_q\in\P\\ \exists\,t:\,P_t\in\mathscr{E}}} \bigg|\prod_{t=1}^q\rho(P_t)\bigg| \left|u(P_1,\dots,P_q)\right| \,\leq\,
\sum_{\substack{P\in\P\\ P\in\mathscr{E}}} |\rho(P)|\,e^{a(P)} \;.
\eeq
\end{theorem}

It is worth to observe that if $(P_1,\dots,P_q)$ is not a cluster then $u(P_1,\dots,P_q)=0$. Therefore only the clusters of polymers (that are infinitely many) give non-zero contributions to the expansion \eqref{eq: absCE logZ} of $\log\ZZ$.

\begin{acknowledgements}
I thank Prof. Elliott H. Lieb for his invitation at Princeton University, for having proposed me to work on his conjecture and for many useful discussions.
I thank Prof. Pierluigi Contucci, Emanuele Mingione and Lukas Schimmer for interesting discussions.
Financial support from UniBo Department of Mathematics, from FIRB grant RBFR10N90W and from PRIN grant 2010HXAW77 is acknowledged.
\end{acknowledgements}

\end{document}